%% file: main.tex
\documentclass[sigplan,10pt]{acmart}

\usepackage[english]{babel}
\usepackage{blindtext}
\usepackage{xcolor}
\usepackage{soul}
\usepackage{bbm}
\usepackage{amsmath}

\usepackage{subcaption}
\usepackage{amsmath}
\usepackage{amsfonts} 
\usepackage{algorithm}
\usepackage{algpseudocode}
\usepackage{xspace}
\usepackage{makecell}
\usepackage{enumitem}

\usepackage{url}

\hypersetup{
    colorlinks=true, 
    linkcolor=black, 
    filecolor=black,
    urlcolor=black,
    citecolor=black, 
}

\renewcommand\footnotetextcopyrightpermission[1]{} 
\setcopyright{none}
\setcopyright{acmcopyright}
\setcopyright{acmlicensed}
\setcopyright{rightsretained}
\setcopyright{usgov}
\setcopyright{usgovmixed}
\setcopyright{cagov}
\setcopyright{cagovmixed}



\begin{document}
\settopmatter{authorsperrow=4}
\title{ATLAS: Expert-Guided Sustainable Satellite Backhaul for Off-Grid Wildlife Monitoring}

\author{Hao Fang}
\orcid{0009-0002-8335-2935}
\affiliation{%
  \institution{Simon Fraser University\\AXSOTA Intelligence Inc.}
  \city{Burnaby}
  \state{BC}
  \country{Canada}
}
\email{fanghaof@sfu.ca}

\author{Miao Zhang}
\authornote{Corresponding author.}
\orcid{0000-0001-6126-6142}
\affiliation{%
  \institution{Simon Fraser University\\AXSOTA Intelligence Inc.}
  \city{Burnaby}
  \state{BC}
  \country{Canada}
}
\email{mza94@sfu.ca}

\author{Haoyuan Zhao}
\orcid{0009-0004-5324-2174}
\affiliation{%
  \institution{Simon Fraser University}
  \city{Burnaby}
  \state{BC}
  \country{Canada}
}
\email{hza127@sfu.ca}

\author{Long Chen}
\orcid{0000-0002-8957-6625}
\affiliation{%
  \institution{Simon Fraser University}
  \city{Burnaby}
  \state{BC}
  \country{Canada}
}
\email{longchen.cs@ieee.org}

\author{Yi Ching Chou}
\orcid{0009-0008-3946-2794}
\affiliation{%
  \institution{Simon Fraser University}
  \city{Burnaby}
  \state{BC}
  \country{Canada}
}
\email{ycchou@sfu.ca}

\author{Chi Xu}
\orcid{0000-0001-9473-722X}
\affiliation{%
  \institution{Simon Fraser University\\AXSOTA Intelligence Inc.}
  \city{Burnaby}
  \state{BC}
  \country{Canada}
}
\email{chix@sfu.ca}

\author{Cheng Pan}
\orcid{0000-0002-8884-4405}
\affiliation{%
  \institution{Simon Fraser University}
  \city{Burnaby}
  \state{BC}
  \country{Canada}
}
\email{chengpan@ieee.org}

\author{Jiangchuan Liu}
\orcid{0000-0001-6592-1984}
\affiliation{%
  \institution{Simon Fraser University}
  \city{Burnaby}
  \state{BC}
  \country{Canada}
}
\email{jcliu@sfu.ca}

\renewcommand{\shortauthors}{Fang et al.}
\renewcommand{\shorttitle}{ATLAS}

\begin{abstract}
Remote wildlife monitoring increasingly depends on high-fidelity cameras and sensors deployed far from grid power and terrestrial networks. Low Earth orbit satellite links can provide connectivity in these settings, but their high power draw and variable bandwidth, combined with intermittent on-site solar harvesting, make continuous backhaul infeasible. Current systems therefore filter at the edge, with motion triggers that waste limited energy and bandwidth on environmental artifacts, or fixed vision models that cannot follow evolving expert objectives.

We present \textsc{ATLAS}, an expert-guided edge-to-cloud system that treats satellite backhaul as a scarce resource for off-grid wildlife monitoring. \textsc{ATLAS} uses foundation models to compile natural-language expert intent into semantic scoring logic, combines semantic relevance with environmental and energy state, and uses a constrained reinforcement-learning scheduler to decide which clips to upload and when. As a representative case study, we deployed \textsc{ATLAS} for over six months across 18 off-grid salmon-monitoring river sites. \textsc{ATLAS} achieves a mean clip-selection accuracy of $87.4\%$, exceeding mean baseline accuracy by $26.4$ percentage points, while maintaining near-continuous operation through low-solar periods in which baselines incur 16.0\% downtime.
\end{abstract}

\maketitle
\pagestyle{empty}

\input{sections/1-introduction}

\input{sections/2-motivation}

\input{sections/3-system}

\input{sections/4-evaluation}

\input{sections/5-related_works}

\input{sections/6-conclusion}

\begin{acks}
We thank the anonymous reviewers and our shepherd for their constructive feedback and guidance on this paper. We are also grateful to the Pacific Salmon Foundation and the Wild Salmon Center for their support and collaboration. This research was supported by an NSERC Discovery Grant and a Mitacs Accelerate Grant.
\end{acks}

\balance
\bibliographystyle{ACM-Reference-Format}
\bibliography{reference}

\end{document}

%% file: sections/1-introduction.tex
\section{Introduction}
Remote wildlife monitoring is a cornerstone of conservation and ecosystem management~\cite{juang2002zebranet, elias2017s}. Modern sensors such as high-definition video traps and continuous acoustic recorders produce large volumes of high-fidelity data for the domain experts who run these programs. Until recently, off-grid sites had no way to move that data off-site in time for experts to act on it.

The recent proliferation of low Earth orbit (LEO) satellite constellations addresses this gap, offering global Internet access and low-latency communication from the edge~\cite{mohan2024starlink, ren2024sateriot, shenoy2024cosmac, de2023hands}. Even with ubiquitous connectivity, off-grid deployments remain bounded by energy~\cite{sorber2007eon, jiang2005prometheus}, since they rely on unstable solar harvesting while the LEO user terminal (UT) draws tens of watts when transmitting, far more than the camera and edge computer it serves (\S\ref{subsec:limited_upload}). During the most severe overcast periods at one of our rainforest sites, daily harvest fell to 21\,Wh, about a tenth of what a standard outdoor camera needs for a day and enough for less than 20 minutes of active transmission on the terminal.

\begin{figure*}[!t]
  \centering
    \begin{subfigure}[b]{0.21\linewidth}
    \includegraphics[width=\linewidth]{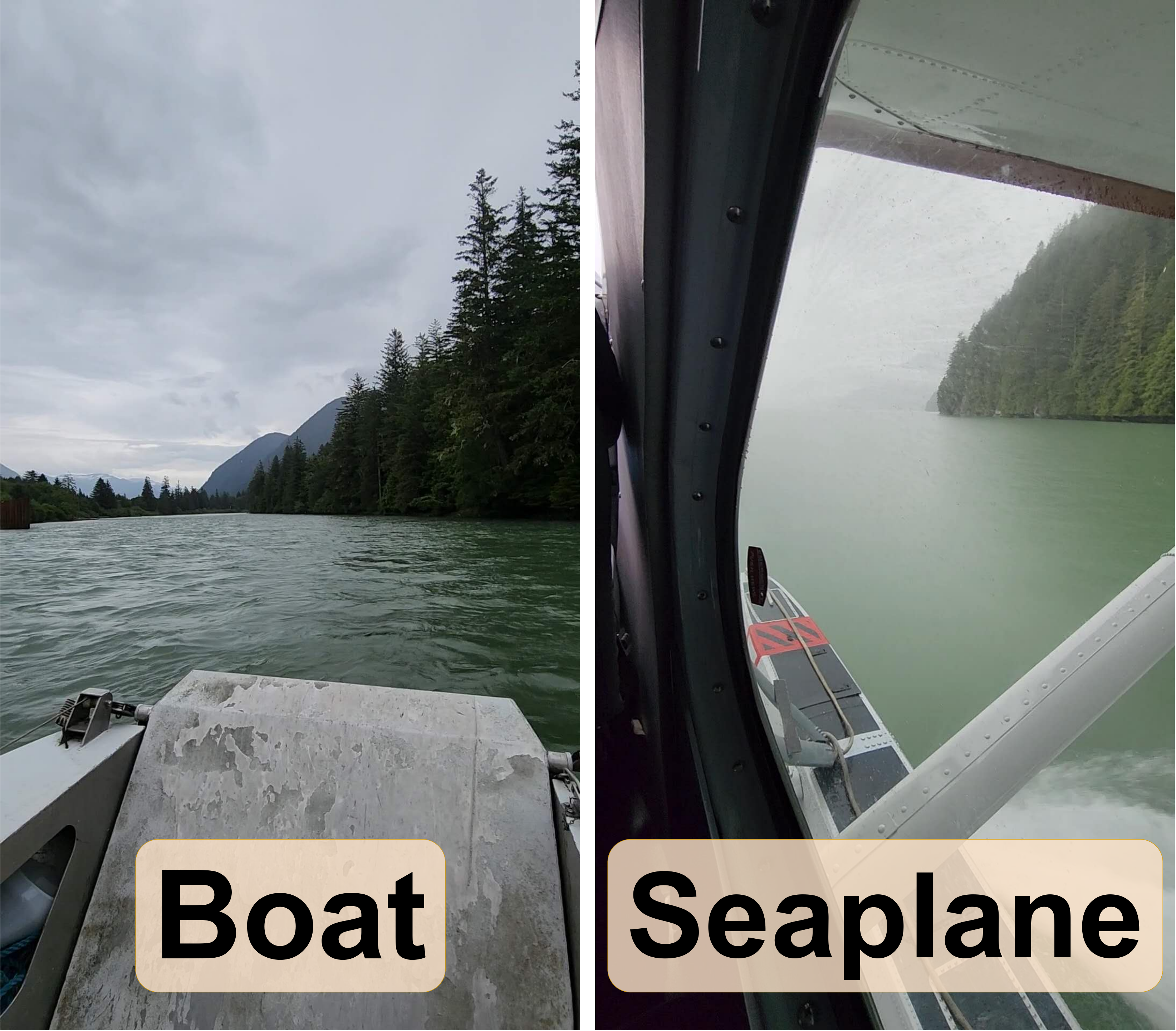}
    \caption{}
    \label{fig:transportation}
  \end{subfigure}
    \hspace{0.046\linewidth}
  \begin{subfigure}[b]{0.1843\linewidth}
    \includegraphics[width=\linewidth]{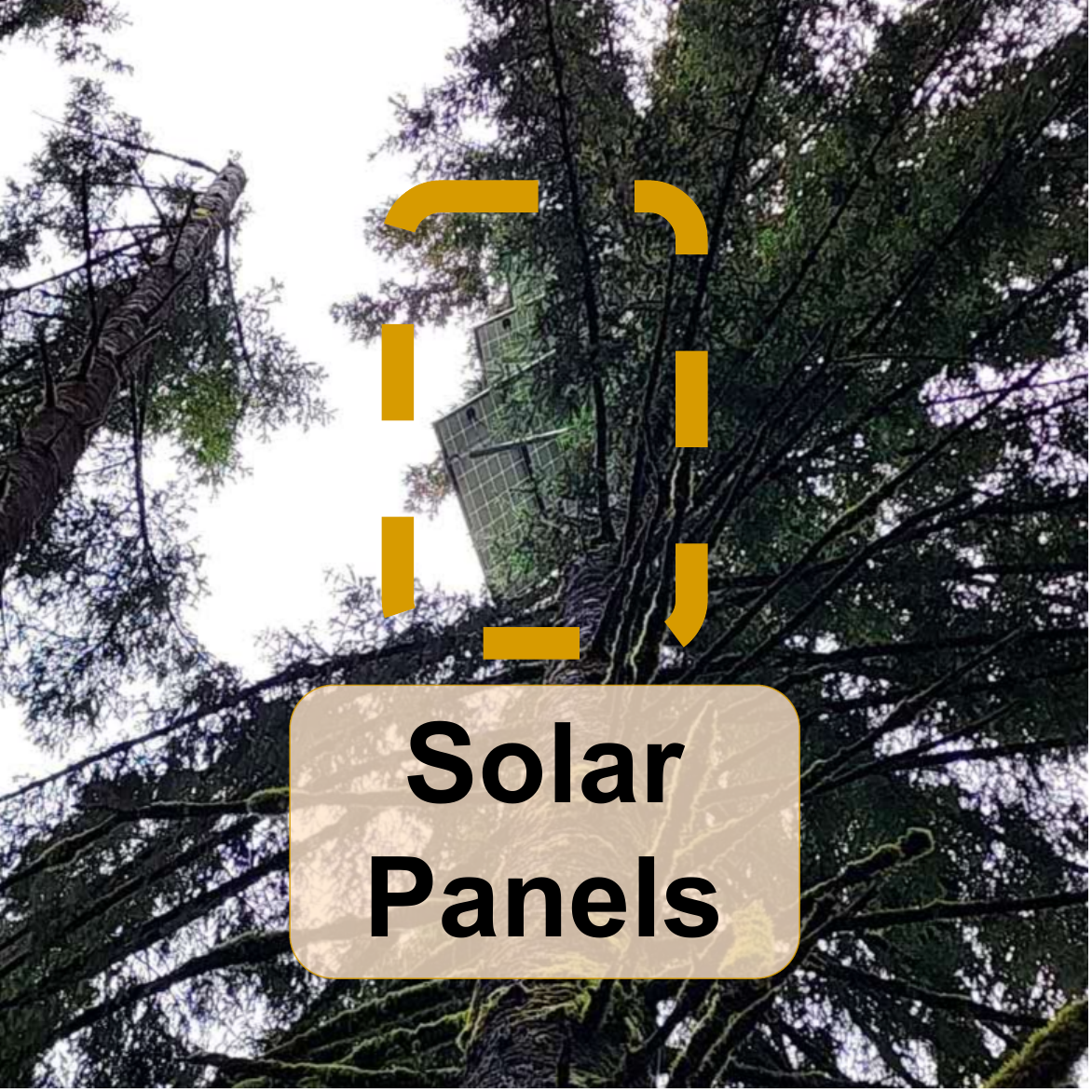}
    \caption{}
    \label{fig:site_pv}
  \end{subfigure}
  \hspace{0.046\linewidth}
  \begin{subfigure}[b]{0.1843\linewidth}
    \includegraphics[width=\linewidth]{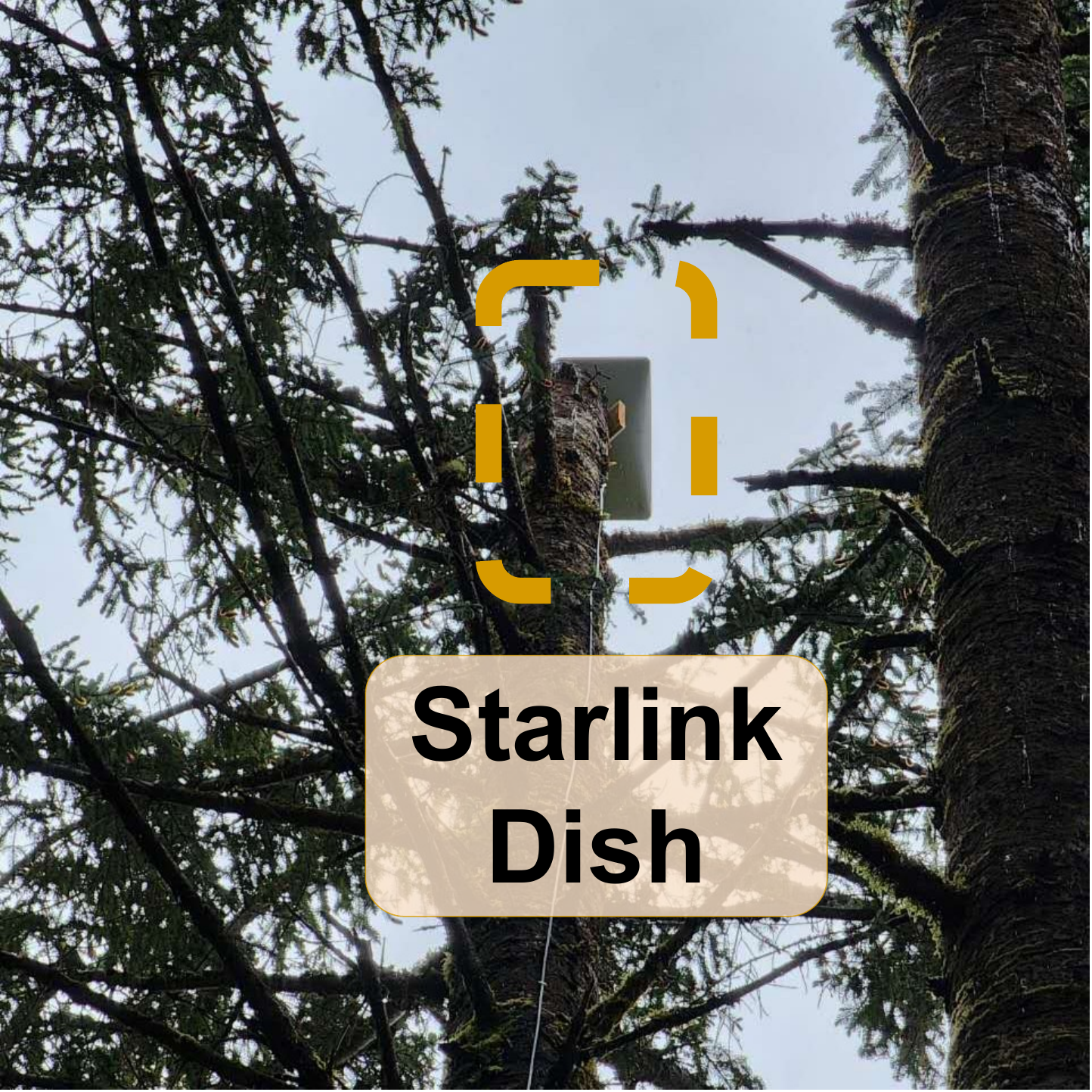}
    \caption{}
    \label{fig:site_dish}
  \end{subfigure}
      \hspace{0.046\linewidth}
    \begin{subfigure}[b]{0.1843\linewidth}
    \includegraphics[width=\linewidth]{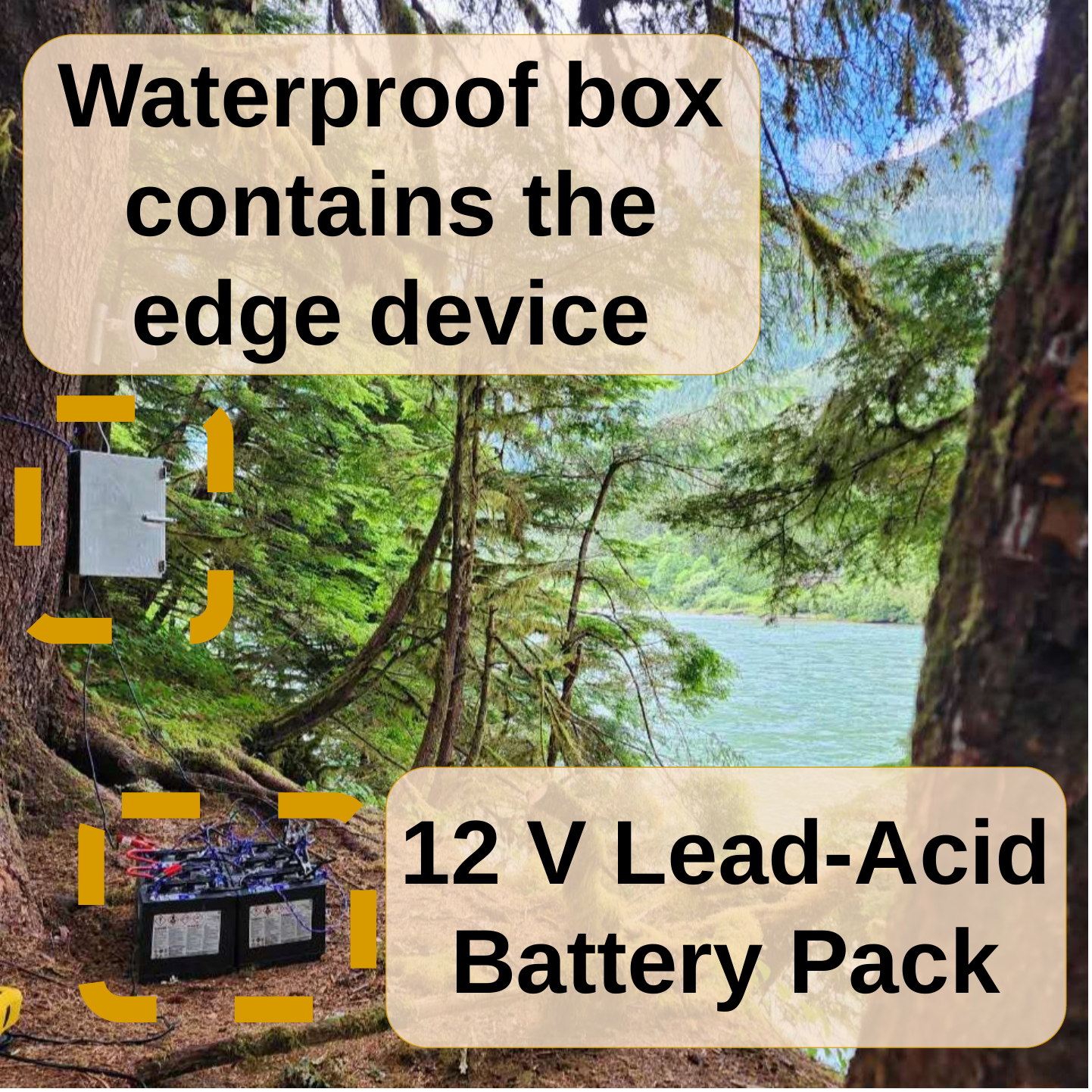}
    \caption{}
    \label{fig:site_system}
  \end{subfigure}
  \vspace{-2mm}
  \captionsetup{width=\linewidth}
    \caption{\textbf{(a)} Reaching deployment sites often requires multiple modes of transportation, including boats and seaplanes, making site visits costly. \textbf{(b, c)} A typical remote wildlife monitoring deployment in a temperate rainforest, with solar panels and the Starlink user terminal (UT) mounted in the tree canopy to maximize sunlight exposure and minimize signal obstruction, respectively. \textbf{(d)} Edge system hardware, including the edge computing enclosure and battery pack.}
  \label{fig:site_access}
\end{figure*}

A long line of work on energy-constrained sensing and communication addresses two questions: what to transmit and when. Motion triggers~\cite{welbourne2016passive} and on-device computer vision (CV) models~\cite{elias2017s, li2020reducto, yuan2022infi} decide which footage is captured or sent. Harvest-aware runtimes adapt duty cycles to predicted energy~\cite{kansal2007power, sorber2007eon}, and energy-aware schedulers send data only when it costs less energy or carries more information~\cite{schulman2010bartendr, wu2012event}. These mechanisms fall short in two respects. First, neither reflects expert intent, since content selection encodes relevance as classes fixed before deployment and transmission scheduling values data by a generic measure of information. Second, each answers only its own question, so content is largely selected without regard to the energy state, and transmissions are scheduled without regard to what the content shows. A monitoring system must therefore decide jointly, based on expert intent, which video clips are worth their energy cost and when to spend it.

Optimizing these decisions first requires understanding what information is actually valuable to domain experts. Through our partnerships with wildlife monitoring organizations, particularly fisheries teams, we find that expert intent is rarely reducible to a small set of static labels. Instead, interests are often expressed as complex, context-dependent tasks articulated in natural language, such as ``identify upstream movement of adult salmon during peak return windows'' or ``flag unusual species appearances following heavy rainfall events.'' Moreover, these objectives are dynamic. Experts may focus on different species across seasons, shifting priority from migration onset to behavioural anomalies or size distributions. Complicating this further, the visual environment is constantly changing. Water turbidity, lighting, and vegetation can shift rapidly, meaning the same high-value event can look markedly different under local conditions.

Foundation models (FMs), particularly large language and multimodal models, offer a way to translate such intent into system behaviour~\cite{radford2021clip, wei2022cot, yao2023react}. Unlike supervised models restricted to fixed classes, they generalize beyond those classes and follow instructions in natural language~\cite{wu2024netllm, zhang-etal-2023-video, yin2024reasoning, wan2024larr}, so that expert knowledge can be supplied in the same form and updated as conservation goals and environmental conditions evolve. However, FMs address only what to transmit and cannot weigh a clip's value to the experts against its energy cost.

We present \textsc{ATLAS}, an energy-aware, expert-guided edge-to-cloud system for off-grid wildlife monitoring. \textsc{ATLAS} treats satellite connectivity as a first-class, budgeted resource and reasons jointly about what a clip is worth to the experts and about the constraints imposed by variable solar energy, high uplink power draw, and time-varying bandwidth. A fine-tuned language model, the Semantic Assessment Agent, reads the compiled expert intent, site-specific knowledge, and frame-level descriptions of each candidate clip, and assigns the clip a relevance score and an uncertainty estimate. All scored candidates enter a local event buffer. The Adaptive Uplink Scheduler combines these assessments with energy telemetry, forecasts, and link measurements, and its reinforcement learning (RL) policy decides \emph{which} clips to upload and \emph{when}, balancing expert relevance, timeliness, and battery reserves. Expert feedback updates the intent and knowledge that the agent reads, so that \textsc{ATLAS} follows evolving goals without retraining.

As a representative case study, we partnered with North American fisheries organizations focused on salmon conservation and deployed \textsc{ATLAS} at 18 off-grid river sites along the Pacific coast. Over a deployment of more than six months spanning the entire salmon migration and spawning season, our evaluation shows that \textsc{ATLAS} effectively interprets experts’ complex intent and incorporates expert knowledge to identify video clips that best match expert interests, thereby reducing unnecessary uplink energy expenditure. Compared with representative baselines, \textsc{ATLAS} identifies expert-relevant video clips with 87.4\% accuracy, achieving an average improvement of $26.4$ percentage points. In addition, \textsc{ATLAS} maintains near-continuous operation throughout extended low-solar periods by proactively budgeting energy for communication, whereas the baselines experience an average system downtime of 16.0\% due to energy depletion. These results demonstrate that expert-guided, energy-aware uplink control can substantially improve both observational utility and long-term system sustainability. Our primary contributions are summarized as follows:

\begin{itemize}
    \item We design \textsc{ATLAS}, an edge-to-cloud wildlife monitoring system that treats satellite connectivity as a scarce, budgeted resource, and decides jointly which video clips are worth transmitting and when.

    \item We build a Semantic Assessment Agent that scores each clip against compiled expert intent and site knowledge, with an uncertainty estimate, and an expert-in-the-loop workflow that updates the intent and knowledge it reads without retraining.

    \item We formulate uplink scheduling as a constrained Mar\-kov decision process that balances semantic utility, information freshness, and battery reserves under stochastic solar harvesting and variable satellite connectivity, with a runtime safety shield that enforces per-step energy limits.
    
    \item We fully implement \textsc{ATLAS} on off-grid hardware and validate it through real field deployments and trace-driven experiments. Our evaluation demonstrates that \textsc{ATLAS} improves both clip-selection accuracy and system uptime over three representative baselines.
\end{itemize}

%% file: sections/2-motivation.tex
\section{Motivation}
Our deployments face constraints long studied in energy-harvesting and delay-tolerant sensing systems~\cite{juang2002zebranet, jiang2005prometheus, kansal2007power, sorber2007eon, dyo2010wildlife, fall2003dtn}. Energy harvest varies with weather and season, installations operate unattended for months, and outages leave permanent gaps in the ecological record. Drawing on observations from our sites, this section examines how these constraints manifest in a LEO-connected video system and identifies the resulting design requirements.

\begin{figure}[!t]
    \centering
    \includegraphics[width=\linewidth]{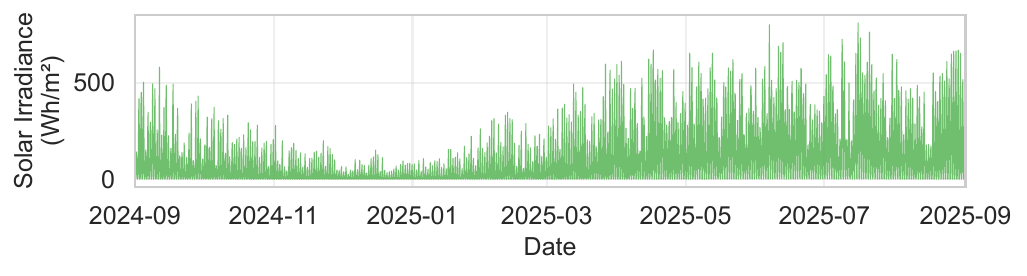}
    \vspace{-5mm}
    \caption{Measured solar irradiance over one year. Higher values correspond to increased solar energy availability, directly impacting battery charging and system uptime.}
    \label{fig:solar_radiation}
    \vspace{-2mm}
\end{figure}

\begin{figure}[!t]
    \centering
    \includegraphics[width=\linewidth]{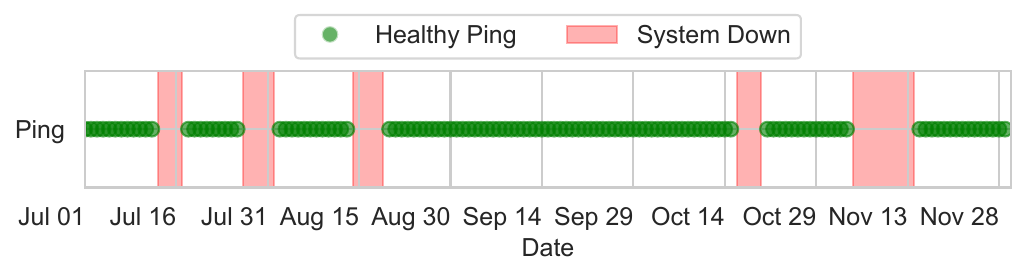}
    \vspace{-5mm}
    \caption{Preliminary system uptime measurements at a representative riverside site on the west coast of North America.}
    \vspace{-2mm}
    \label{fig:site_outage_example}
\end{figure}

\subsection{Challenges of Wild Deployment Environments}
\label{sec:challenging_wild_condition}
Wildlife monitoring is increasingly conducted in remote, ecologically sensitive habitats, including tropical and temperate rainforests, high-latitude wetlands, alpine regions, and protected river systems, where reliable grid power and terrestrial Internet connectivity are often unavailable. Our monitoring sites along salmon-bearing rivers on the Pacific coast of North America exemplify these conditions. As shown in Figure~\ref{fig:site_access}, each site operates entirely off-grid, relying on solar energy harvesting for power and a satellite link for Internet connectivity. This reliance on solar power makes energy availability dependent on local weather conditions.

Figure~\ref{fig:solar_radiation} presents one year of irradiance measurements from a temperate rainforest site. Solar energy availability is substantially higher in summer than in winter and exhibits pronounced day-to-day variability within each season. Figure~\ref{fig:site_outage_example} illustrates the operational consequences of this variability. During a preliminary deployment at a riverside site, five episodes of adverse weather led to battery depletion and system shutdowns, with a median downtime of 120 hours. Recovery from each shutdown required manual intervention. An upload that appears feasible given the current battery state of charge may therefore lead to a shutdown several days later. Uplink energy must be budgeted over multi-day horizons using energy-harvest forecasts.

\begin{figure}[!t]
    \centering
    \includegraphics[width=\linewidth]{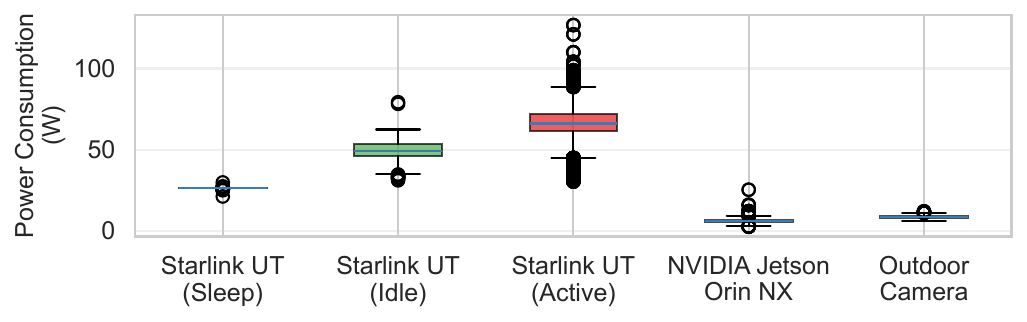}
    \vspace{-5mm}
    \caption{Measured power draw of on-site components, including the Starlink Standard (Gen 3) user terminal (UT) in sleep, idle, and active states.}
    \vspace{-2mm}
    \label{fig:starlink_power}
\end{figure}

\begin{figure}[!t]
    \centering
    \includegraphics[width=\linewidth]{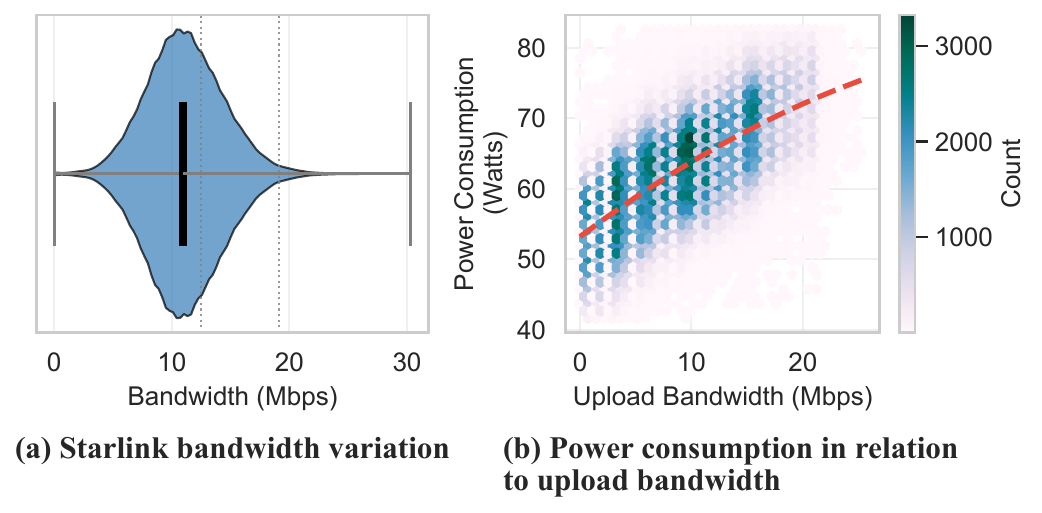}
    \vspace{-5mm}
    \caption{Measured Starlink UT upload bandwidth and its effect on power draw.}
    \vspace{-2mm}
    \label{fig:starlink_power_bandwidth}
\end{figure}

\begin{figure*}[!t]
  \centering
    \begin{subfigure}[b]{0.21\linewidth}
    \includegraphics[width=\linewidth]{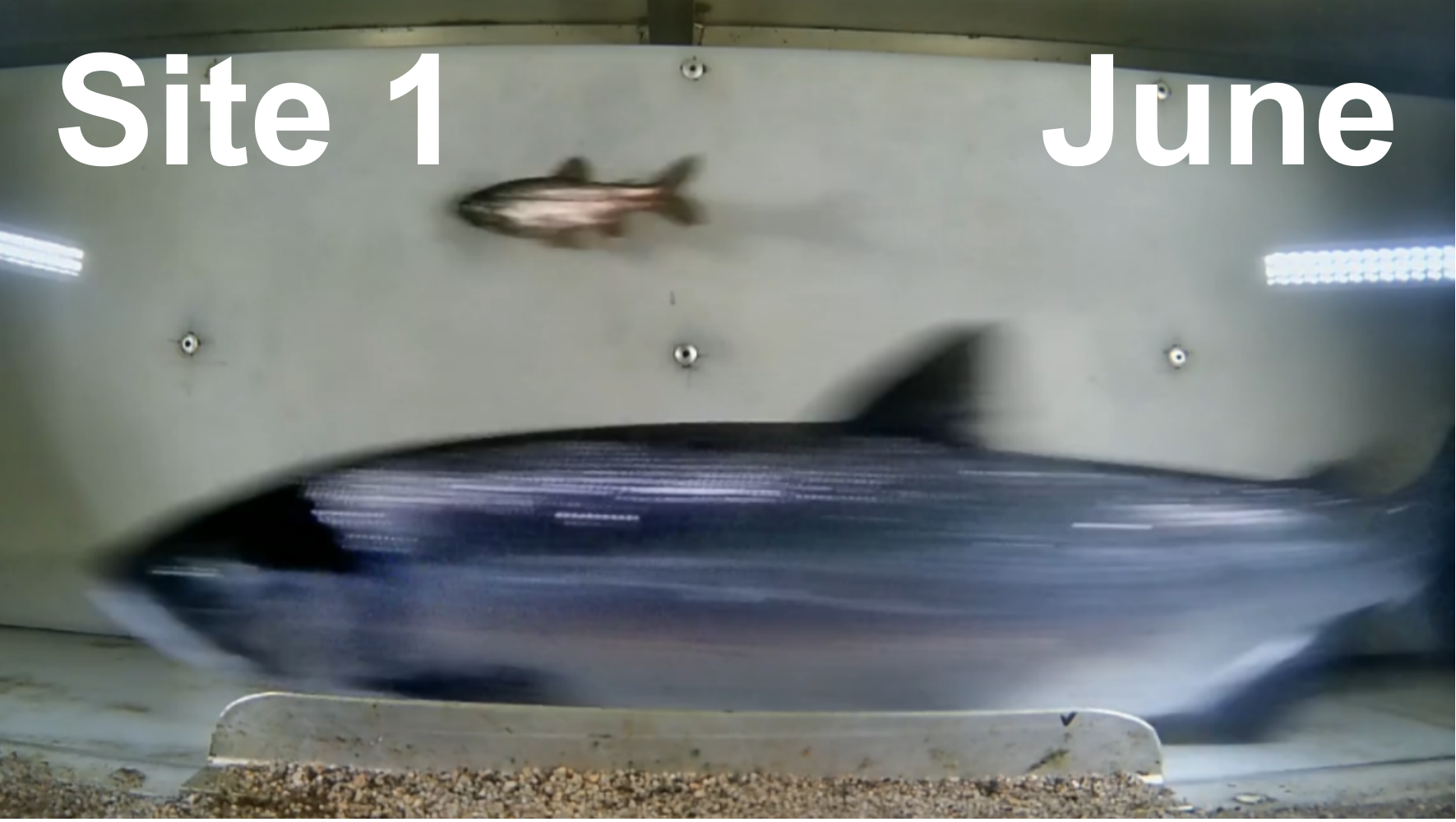}
    \caption{}
    \label{fig:fish_a}
  \end{subfigure}
    \hspace{0.02\linewidth}
  \begin{subfigure}[b]{0.21\linewidth}
    \includegraphics[width=\linewidth]{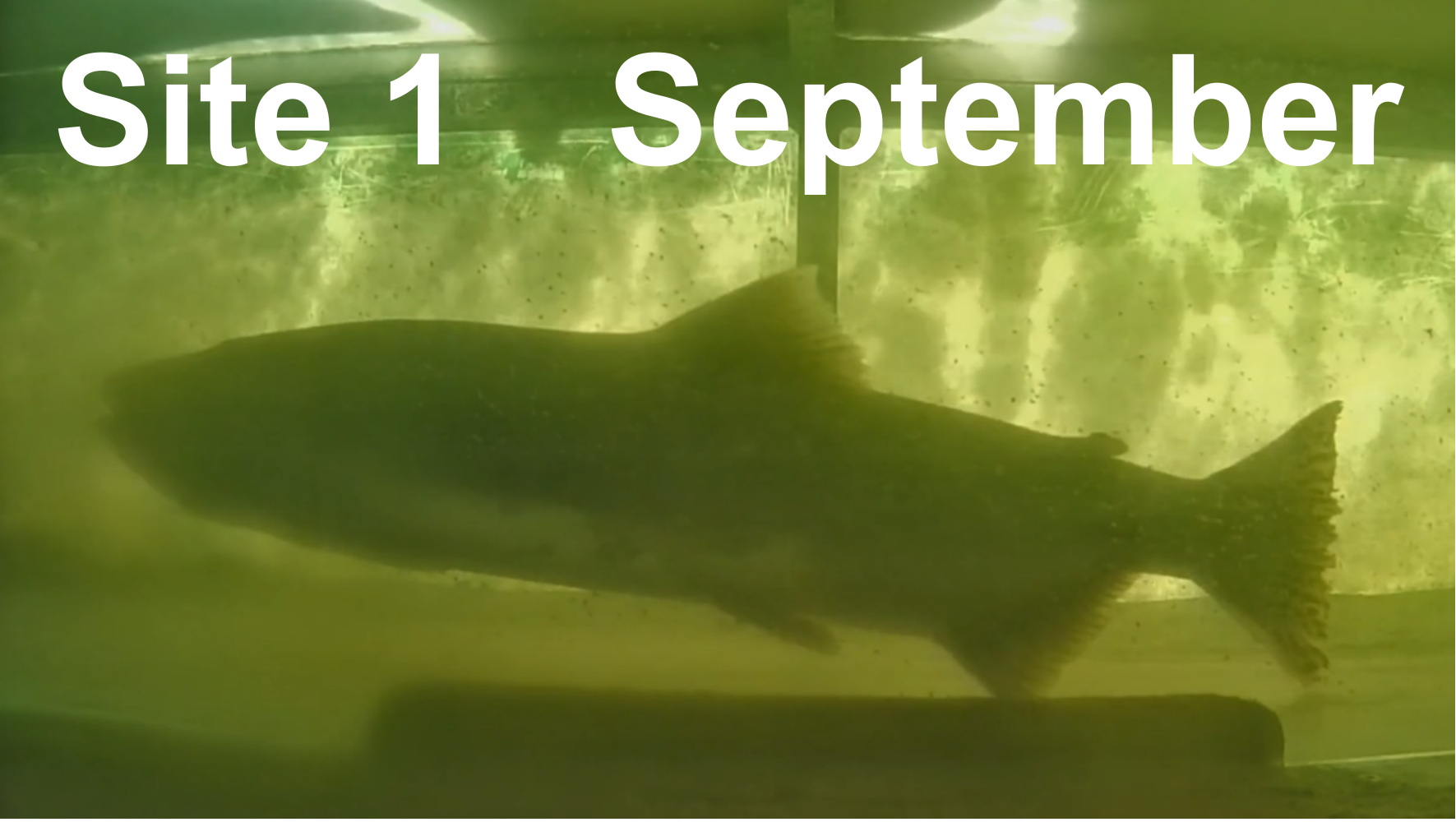}
    \caption{}
    \label{fig:fish_b}
  \end{subfigure}
  \hspace{0.02\linewidth}
  \begin{subfigure}[b]{0.21\linewidth}
    \includegraphics[width=\linewidth]{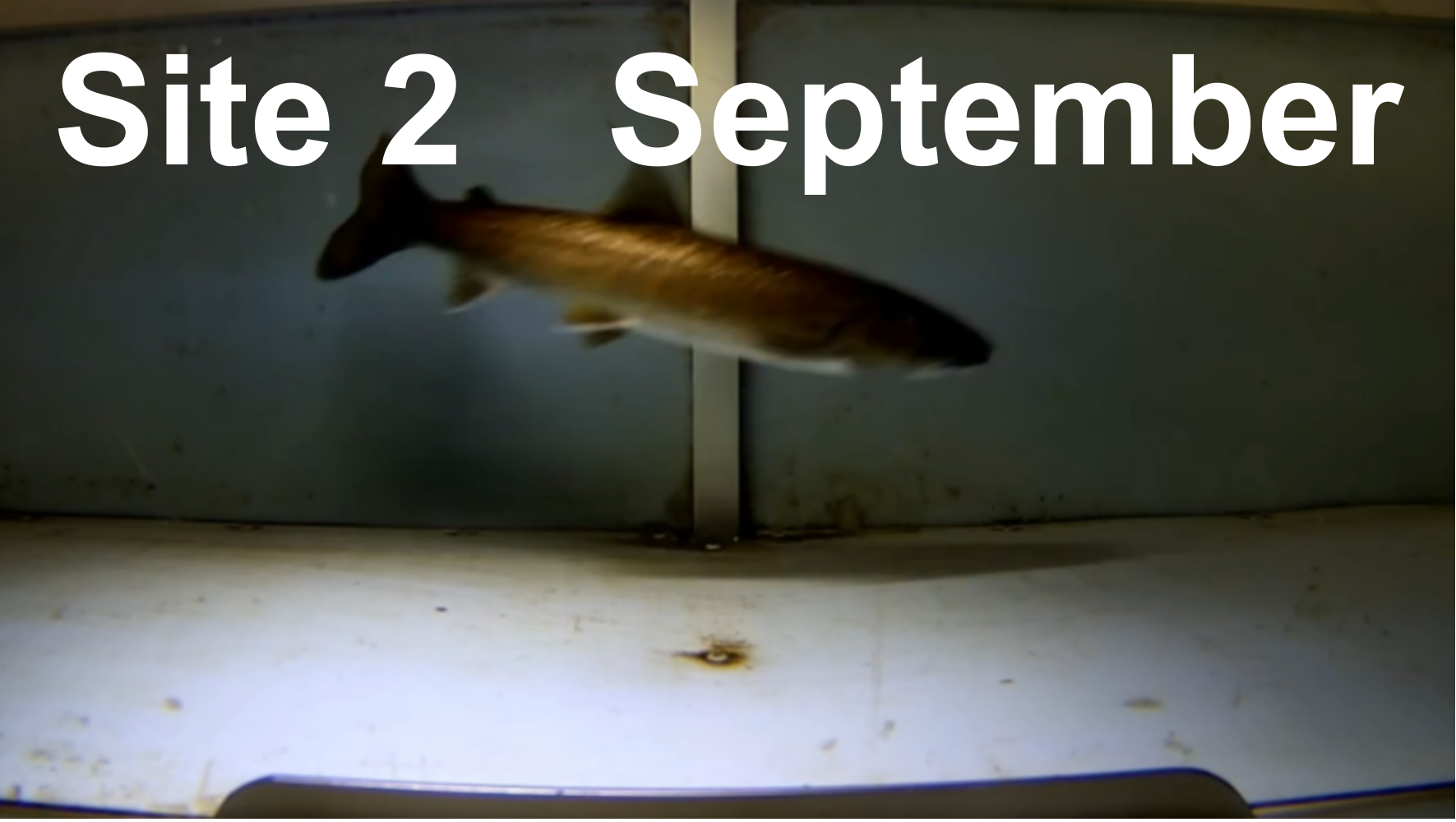}
    \caption{}
    \label{fig:fish_c}
  \end{subfigure}
      \hspace{0.02\linewidth}
    \begin{subfigure}[b]{0.21\linewidth}
    \includegraphics[width=\linewidth]{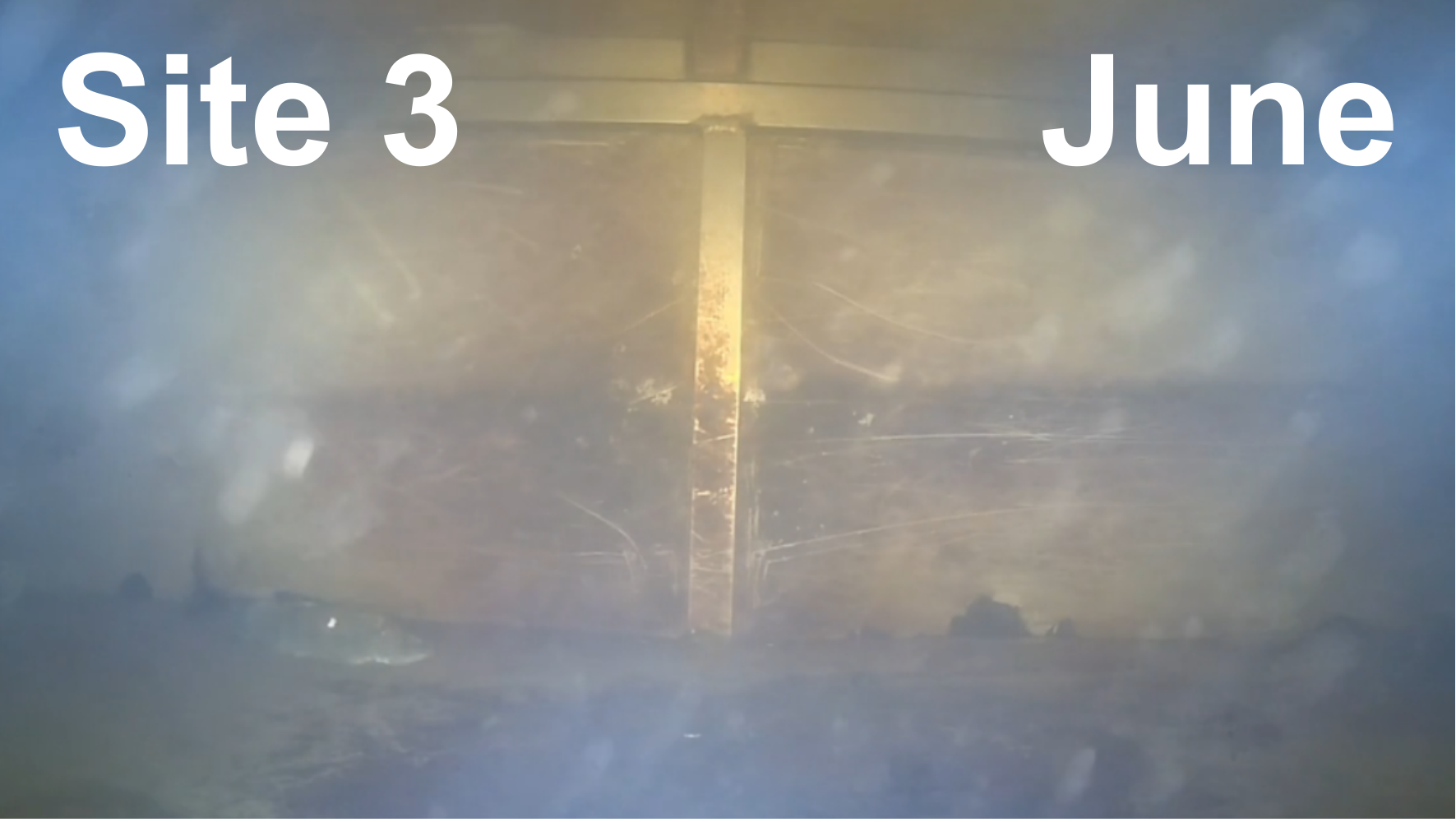}
    \caption{}
    \label{fig:fish_d}
  \end{subfigure}
  \vspace{-3mm}
  \captionsetup{width=\linewidth}
  \caption{Visual variability in salmon footage, showing seasonal change at one site and differing conditions across sites.}
  \label{fig:species_variation}
  \vspace{-1mm}
\end{figure*}

\subsection{Power and Bandwidth Constraints of Satellite Communication}
\label{subsec:limited_upload}
At these sites, satellite communication is the only means of transferring data off-site. Today, LEO services such as Starlink provide connectivity to these remote sites at a reasonable cost, but for an off-grid system, the link imposes two further constraints: a high, bursty power draw and a limited, variable uplink bandwidth.

Figure~\ref{fig:starlink_power} shows the measured power draw of the on-site components. The Starlink UT consumes substantially more power than any other component, even in sleep and idle states, whereas the edge computer and camera typically draw less than 10\,W. In addition, waking the terminal and establishing a usable link takes 183\,s at 50.6\,W, corresponding to a fixed energy cost of approximately 9.26\,kJ, independent of the volume of data transmitted. These costs limit the number of hours per day for which the terminal can be kept awake within the available energy budget. Uploads must therefore be batched to amortize the wake-up cost rather than transmitted as individual events occur.

Figure~\ref{fig:starlink_power_bandwidth}(a) shows the measured uplink throughput, which ranges from 0 to 30.33\,Mbps with a median of 10.78\,Mbps and varies with weather conditions, terrain occlusion, and antenna orientation. Power draw also rises with throughput (Figure~\ref{fig:starlink_power_bandwidth}(b)). This variability makes the transmission time and energy required to deliver a clip uncertain in advance. Upload batches must be sized according to forecasts of available uplink bandwidth.

Together, these constraints create a trade-off between timely data delivery and sustained system operation. Uploading more supports downstream analysis but consumes energy and link time, whereas uploading less conserves energy but can delay critical information. Compression helps only up to a point, because a clip reduced below the fidelity that experts need to verify it loses its value. A keyframe set, for example, cannot show movement when the task depends on it. Which clips to upload, how much to send in each window, and when to wake the terminal must therefore be planned jointly, using forecasts of both available bandwidth and energy harvest.

\subsection{Concept Drift and the Limits of Static Models}
Wildlife monitoring systems operate in dynamic and heterogeneous environments, so a single model trained on data from a limited set of conditions rarely generalizes reliably. Models that perform well in one river, season, or lighting condition may degrade significantly when applied to another~\cite{beery2018terra}. Long-term deployment amplifies the problem, since the data distribution drifts away from the training distribution as the environment evolves, and performance degrades over time.

Wildlife monitoring also presents intrinsic challenges from species diversity and visual ambiguity. Many species, such as salmonids, differ only subtly from one another, and individuals of the same species vary in appearance, so they are difficult to distinguish from appearance alone, especially under noisy sensing conditions. Figure~\ref{fig:species_variation} shows salmon species that appear visually similar alongside footage captured under diverse water conditions. Trained experts resolve these distinctions by drawing on contextual cues such as seasonality, behaviour, and migration patterns, which conventional CV models cannot capture from raw imagery alone.

Much of this expert knowledge is not encoded in labelled datasets. It exists in unstructured forms such as field notes, written reports, regulatory guidelines, and oral knowledge passed between practitioners, for example ``count only up\-stream-mov\-ing adults,'' ``ignore glare-induced false positives,'' or ``flag unusually pale fish during low flow.'' Detection and classification models such as YOLO-style architectures~\cite{redmon2016you} are trained from visual supervision over a fixed label set and offer no way to incorporate such rules. A monitoring system must take expert knowledge in the form in which it exists, as natural-language rules and descriptions, and revise what counts as relevant as that knowledge changes, without retraining or large-scale labelling.

%% file: sections/3-system.tex
\section{ATLAS Overview}
\begin{figure}[t!]
    \centering
    \includegraphics[width=\linewidth]{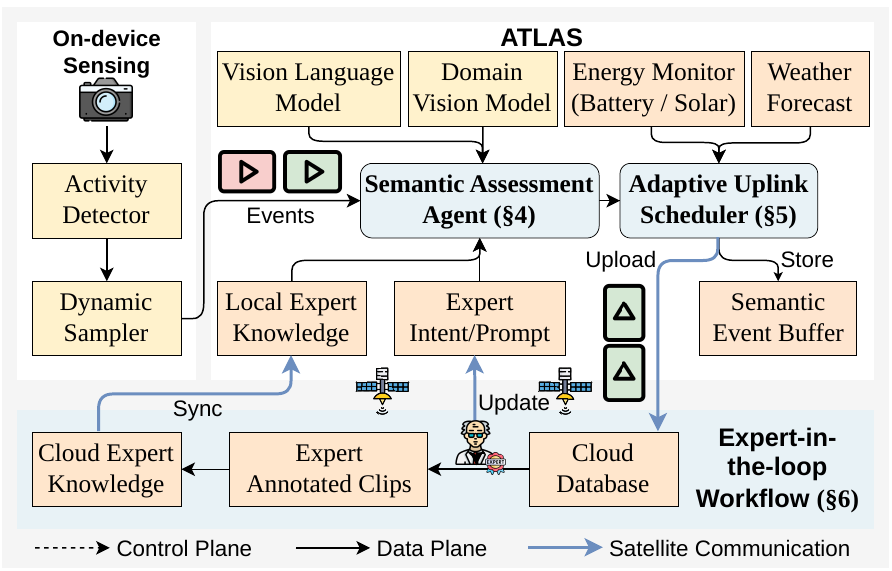}
    \vspace{-8mm}
    \caption{ATLAS system overview}
    \label{fig:system_overview}
    \vspace{-5mm}
\end{figure}

ATLAS uses expert-guided semantic assessment to identify relevant observations from remote, off-grid deployments. Its energy-aware scheduler determines which observations to upload over the satellite link and when, given the available energy and bandwidth. Figure~\ref{fig:system_overview} illustrates the end-to-end workflow of ATLAS.

Low-power sensing runs continuously on the device. A lightweight Activity Detector watches the camera feed for image-based motion cues, and the Dynamic Sampler adapts the frame rate to its output. While no activity is detected, the sampler keeps a low frame rate to minimize computation and storage. When activity is detected, it raises the rate and extracts a short clip that preserves enough visual detail for analysis. Each clip forms a candidate event, the unit that the rest of the system assesses and schedules, so that later stages spend energy only on footage that contains activity.

For each candidate event, the \textbf{Domain Vision Model (DVM)}, fine-tuned on footage from our deployment sites, labels and tracks objects in the clip, and the \textbf{Vision Language Model (VLM)} adds a holistic description of each frame when the expert intent requires such context. The \textbf{Semantic Assessment Agent} reads these frame-level descriptions together with the compiled expert intent and the Local Expert Knowledge, which stores the site's expert knowledge and event memory for retrieval. It outputs a relevance score, an uncertainty estimate, and the frames that support them (\S\ref{sec:agent}). Every candidate event, including low-scoring ones, enters the \textbf{Semantic Event Buffer} with its clip and assessment.

The \textbf{Adaptive Uplink Scheduler} combines these assessments with battery and solar telemetry from the Energy Monitor, the Weather Forecast, and link measurements. It decides which events to upload, keep, or discard, and when to wake the satellite terminal for transmission (\S\ref{sec:scheduler}).

The \textbf{Expert-in-the-Loop Workflow} closes the loop (\S\ref{sec:expert-loop}). Uploaded clips reach the Cloud Database, where experts review them, and their judgments, frame marks, and written feedback are kept as Expert Annotated Clips and in the Cloud Expert Knowledge. The resulting updates to the expert intent, the knowledge, and the event memory are synchronized back to the edge over the satellite link, while the agent's weights stay fixed. To keep this feedback representative, ATLAS reserves a small share of each upload budget for sentinel clips, selected by random sampling and by agent uncertainty rather than by score, so that experts also see events the agent rated low.

\section{Semantic Assessment Agent}
\label{sec:agent}

The Semantic Assessment Agent, hereafter the agent, is a fine-tuned language model that assesses each candidate clip against the current expert intent. It reads an assembled textual context and emits a structured assessment that the Adaptive Uplink Scheduler (\S\ref{sec:scheduler}) uses to determine what to upload.

\subsection{Context and Assessment}
\label{sec:agent-design}

\paragraph{Context assembly.} For each candidate event $c$ and expert intent $\mathcal{I}$, the agent reads an assembled context
\begin{equation}
\label{eq:agent-input}
  x = \big(\mathrm{Compile}(\mathcal{I}),\, \mathrm{Describe}(c),\, \mathrm{Retrieve}(\mathcal{I}, c),\, r(c)\big),
\end{equation}
where $\mathrm{Retrieve}(\mathcal{I}, c)$ collects the entries retrieved from the expert knowledge store $\mathcal{K}$ and the semantic memory $\mathcal{M}$. The components are as follows.

\begin{itemize}[leftmargin=1.5em, labelwidth=0.7em, labelsep=0.3em]
\item \textbf{Compiled expert intent.} An offline step compiles the intent $\mathcal{I}$ into a rubric $\mathrm{Compile}(\mathcal{I})$. The rubric has six fixed fields: \texttt{target}, \texttt{exclude}, \texttt{count}, and \texttt{motion}, together with \texttt{condition}, which states when the intent applies, and \texttt{evidence}, which lists the attributes that the descriptions must cover. An open list, \texttt{criteria}, adds free-text checks such as an anomaly descriptor. The rubric is cached and appended to every assessment prompt.

\item \textbf{Frame descriptions.} $\mathrm{Describe}(c)$ converts every candidate event into text, one line per sampled frame of the form \texttt{[frame index] [description]}. The lightweight DVM, an object detector fine-tuned to the deployment domain with a tracking-by-detection step, runs on every event and contributes classes, counts, positions, and track identities. The heavier VLM adds a holistic description of each frame and runs only when the rubric's \texttt{evidence} field asks for such context and the DVM reports a class that is not on the rubric's \texttt{exclude} list.

\item \textbf{Retrieved expert knowledge.} The expert knowledge store $\mathcal{K}$ is a lightweight, site-specific collection of rules of thumb, seasonal priors, known failure modes (e.g., ``bubbles trigger false positives''), and labelling guidelines. The top-$k$ entries most relevant to the intent are retrieved.

\item \textbf{Retrieved event memory.} The semantic memory $\mathcal{M}$ is a rolling record of the events uploaded from the site, with their frame descriptions, regime labels, and any expert feedback. Records similar to the new event are retrieved, supplying expert corrections and site history.

\item \textbf{Regime label.} The regime label $r(c)$ names the viewing conditions of the clip, such as clear water, turbid water, low light, or glare. A small on-device classifier assigns it from cheap image statistics such as contrast, blur, and a turbidity proxy, so that the agent knows under which conditions the descriptions were produced.
\end{itemize}

\paragraph{Assessment interface.} Given context $x$, the agent samples a structured assessment $y$ from its policy $\pi_\theta(y \mid x)$ with parameters $\theta$,
\begin{equation}
\label{eq:agent-output}
  y = \big(s,\, u,\, \mathcal{F},\, \text{rationale}\big),
\end{equation}
where $s$ is the relevance score, $u$ is the uncertainty estimate, and $\mathcal{F}$ is the set of frame indices cited as evidence. When multiple assessments of the same event are compared (\S\ref{sec:agent-training}), the fields of assessment $y$ are denoted by $s_y$, $u_y$, and $\mathcal{F}_y$. The output conforms to a fixed JSON schema with the fields \texttt{score}, \texttt{uncertainty}, \texttt{evidence\_frames}, and \texttt{rationale}.

The score $s \in [0,1]$ is the agent's estimate of $P(z = 1 \mid x)$, where $z \in \{0,1\}$ is the expert's judgment of whether the clip satisfies the intent. The uncertainty $u \in [0,1]$ is the agent's prediction of its own error $|s - z|$. A score near $0.5$ means that the evidence in the clip supports either judgment. A high $u$ means that the agent may be confidently wrong, typically under unfamiliar conditions such as a new site or turbid water, which the regime label $r(c)$ helps it recognize. Every assessment must cite the frames $\mathcal{F}$ whose descriptions support it. The citations tie the score to the observed content of the clip rather than to prior context.

\subsection{Training and In-context Adaptation}
\label{sec:agent-training}

\paragraph{Cold-start distillation.} We first train the agent by knowledge distillation at the sequence level~\cite{kim2016sequence} from a stronger teacher model, so that it produces the schema of Eq.~\eqref{eq:agent-output} from a sensible prior. Each training example pairs a clip with a rubric. Rubrics are either compiled from expert-written intents (\S\ref{sec:agent-design}) or synthesized by instantiating the schema with alternative field values, so that distillation covers a broader range of intents than the few written by experts. For each pair, we assemble the context $x$ as at deployment, let the teacher emit a complete assessment, and fine-tune the agent's backbone to reproduce it. Teacher outputs that contradict $\mathrm{Describe}(c)$, for example by citing frames that do not exist, are discarded. This stage alone makes the agent deployable before any expert review.

\paragraph{Refinement with expert judgments.} Distillation inherits the teacher's blind spots, since the teacher has never seen an expert's verdict on these clips. We therefore refine the agent with expert reviews. To limit expert effort, each review consists of a binary verdict $z \in \{0,1\}$ on whether the clip satisfies the intent and the frames the expert found decisive for it. A classifier would suffice for the score alone, but the evidence frames and the uncertainty estimate must stay consistent with it and their rewards depend on the decoded output, so we treat the verdict as a reward and refine the single generative policy with Group Relative Policy Optimization (GRPO)~\cite{shao2024deepseekmath}. GRPO is the standard optimizer for language-model policies with rule-based rewards~\cite{deepseekr1}.

\emph{Reward.} An assessment $y$ with relevance score $s_y$, uncertainty estimate $u_y$, and cited frames $\mathcal{F}_y$ receives the reward
\begin{equation}
\label{eq:reward}
\begin{aligned}
  R(x, y) = {}& -\mathrm{LogLoss}(s_y, z) \\
              & + \rho\, \mathrm{EvidenceMatch}(y, c) \\
              & - \nu \big(u_y - |s_y - z|\big)^2,
\end{aligned}
\end{equation}
with weights $\rho, \nu > 0$. $\mathrm{LogLoss}$ is the binary cross-entropy, evaluated with $s_y$ clipped to $[\epsilon, 1-\epsilon]$ for a small constant $\epsilon \in (0, \tfrac{1}{2})$; as a strictly proper scoring rule, the log-loss term encourages calibrated scores and penalizes overconfident errors. $\mathrm{EvidenceMatch}(y, c)$ is the intersection over union of $\mathcal{F}_y$ and $\mathcal{A}(c)$, where $\mathcal{A}(c)$ is the set of frames an expert marked as decisive for $z$, indexed as in $\mathrm{Describe}(c)$, and is defined as one when both sets are empty. The union in the denominator penalizes citing unmarked frames, so when no frames are marked, citing none maximizes the term. The final term is maximized when $u_y$ equals $|s_y - z|$, the realized absolute error of the relevance score, so the agent learns to report its error as its uncertainty.

\emph{Sampling and update.} Training inputs are assembled exactly as at deployment. For each training input $x$, we sample a group of $G$ assessments $\{y_1, \ldots, y_G\}$ from the current policy. Because the outputs are sampled, the group differs in score, uncertainty, cited frames, and rationale. Each assessment is scored with the reward of Eq.~\eqref{eq:reward} against the same expert judgment $z$, and its advantage is the reward normalized within the group. The update scales the log-probability of every token of each sampled assessment by its advantage, raising assessments that scored above the group mean and lowering the rest, subject to a Kullback--Leibler (KL) penalty towards the distilled policy that keeps the refinement close to the prior. The rationale needs no reward term of its own, since rationales that accompany well-rewarded assessments are reinforced through this token-level update.

\paragraph{In-context adaptation.} Both training stages run offline on a GPU server before deployment (\S\ref{sec:eval}). During deployment, the agent adapts through context updates while its weights remain fixed. Expert feedback is incorporated in two ways. Interpretive feedback is stored for retrieval, either in the event's record in $\mathcal{M}$ or as a site- or season-specific heuristic in $\mathcal{K}$. Visual cues, such as markings that distinguish two species, are added to the rubric's \texttt{evidence} field. The rubric is then recompiled so that the VLM includes these cues in subsequent descriptions. Changes in expert intent also trigger rubric recompilation. The updated rubric and retrieval stores are synchronized to the edge via satellite.

\section{Adaptive Uplink Scheduler}
\label{sec:scheduler}

The Adaptive Uplink Scheduler decides which buffered events to upload and when, under intermittent connectivity and limited energy. We formulate upload control as a Constrained Markov Decision Process (CMDP)~\cite{altman2021constrained} that balances delivered semantic value, data freshness, and battery reserves. The scheduler combines the agent’s scores and uncertainty estimates with buffer state, energy telemetry, and forecasts. A runtime safety shield checks each resulting upload plan against projected energy availability and connectivity.

\subsection{State, Action, and Transition}
\label{sec:sched-sat}

\paragraph{State.} Decisions are made at discrete steps $t$, one per decision interval $[T_t, T_{t+1})$, where $T_t$ is the clock time at the start of step $t$. The buffer, battery telemetry, and forecasts indexed by $t$ are evaluated at $T_t$, before the action of step $t$ is applied, whereas the plan and upload outcomes of step $t$ refer to the interval that follows. Let $\mathcal{B}_t$ be the set of pending events in the Semantic Event Buffer at step $t$. Each pending event $i \in \mathcal{B}_t$ carries its capture time $T^{\mathrm{cap}}_i$ and the score $s_i$ and uncertainty estimate $u_i$ that the agent assigned when the event entered the buffer (\S\ref{sec:agent-design}), which are not revised while it is pending. At step $t$, the scheduler's policy observes a compact summary $\mathbf{o}_t$ of this state:
\begin{equation}
  \begin{split}
    \mathbf{o}_t = \big(
      &\underbrace{\mathrm{SOC}_t,\, \widehat{H}_{t:t+h-1}}
        _{\text{energy state and forecast}},\;
      \underbrace{\widehat{\mathrm{BW}}_{t:t+h-1},\, q_t,\, \bar{L}_t,\, \ell^{\max}_t}
        _{\text{link forecast and buffer state}}, \\[2pt]
      &\underbrace{p^{\mathrm{prio}}_t,\, \bar{u}_t,\, \bar{s}_t}
        _{\text{semantic context}},\;
      \underbrace{\tau_t,\, \mathrm{DoY}_t}_{\text{time context}}
    \big).
  \end{split}
  \label{eq:sched-state}
\end{equation}
Here ${\mathrm{SOC}_t \in [0,1]}$ is the battery state of charge, and $\widehat{H}_{t:t+h-1}$ and $\widehat{\mathrm{BW}}_{t:t+h-1}$ are the predicted energy harvest and uplink bandwidth for each of the next $h$ decision intervals. The pending payload volume is $q_t$ bytes. The data lag $\bar{L}_t$ is the mean age $T_t - T^{\mathrm{cap}}_i$ of the pending events, and $\ell^{\max}_t$ is their maximum age. The quantity $p^{\mathrm{prio}}_t$ is the fraction of pending events whose score exceeds a fixed reference level. Time of day $\tau_t$ and day of year $\mathrm{DoY}_t$ provide temporal context. The mean score $\bar{s}_t$ and mean uncertainty $\bar{u}_t$ over a nonempty buffer are
\begin{equation}
  \bar{s}_t = \frac{1}{\lvert \mathcal{B}_t \rvert}
    \sum_{i \in \mathcal{B}_t} s_i, \qquad
  \bar{u}_t = \frac{1}{\lvert \mathcal{B}_t \rvert}
    \sum_{i \in \mathcal{B}_t} u_i ,
  \label{eq:sched-summaries}
\end{equation}
so that $\bar{s}_t$ and $p^{\mathrm{prio}}_t$ together convey the shape of the score distribution. For an empty buffer, $\bar{s}_t$, $\bar{u}_t$, $p^{\mathrm{prio}}_t$, $\bar{L}_t$, and $\ell^{\max}_t$ are set to zero, and $q_t = 0$ distinguishes this case from a buffer whose events all scored zero. The policy sees only this summary, whereas plan construction reads the per-event scores, sizes, and sentinel marks in $\mathcal{B}_t$ directly.

\paragraph{Energy reference.} The scheduler estimates the energy stored in the battery at the start of step $t$, and the reserve, as
\begin{equation}
  E_t = \mathrm{SOC}_t\, E^{\mathrm{cap}}, \qquad
  E_{\min} = \mathrm{SOC}_{\min}\, E^{\mathrm{cap}},
  \label{eq:sched-energy}
\end{equation}
where $E^{\mathrm{cap}}$ is the usable battery capacity and ${\mathrm{SOC}_{\min} \in (0,1)}$ is the reserve state of charge. The software reserve $E_{\min}$ is set above the battery management system's protective low-voltage cutoff and is the common reference for the reward, the constraint cost, and the shield's energy guard. Only $\mathrm{SOC}_t$ enters $\mathbf{o}_t$, since $E_t$ is a fixed multiple of it.

\paragraph{Action.} The control policy $\pi_{\phi}$, with parameters $\phi$, outputs $a_t = [b_t, \gamma_t]$, where $b_t \ge 0$ is the total upload budget in bytes and ${\gamma_t \in [0,1]}$ is the score threshold for retaining an event and making it eligible for upload, distinct from the fixed reference level used for $p^{\mathrm{prio}}_t$. Uploads use a representation $\kappa$ from a fixed ladder ordered by payload size: a keyframe set, a reduced-resolution clip, or the full clip. Rather than being learned, $\kappa$ is set at each rubric compilation to the smallest representation that lets experts verify the rubric (\S\ref{sec:agent-design}), such as a keyframe set when the rubric constrains only appearance and a clip when its \texttt{motion} or \texttt{count} field requires experts to see movement. The choice of $\kappa$ therefore affects which events fit within the budget but not the value credited to an individual event in Eq.~(\ref{eq:sched-value}). Let $w_i$ denote the payload size of event $i$ in bytes under the current $\kappa$, recomputed for every pending event before the next decision step whenever a recompilation changes $\kappa$, so that $q_t = \sum_{i \in \mathcal{B}_t} w_i$. Ordinary events with $s_i < \gamma_t$ are discarded from the buffer, and those with $s_i \ge \gamma_t$ are considered for transmission. Sentinel events, which are designated for expert audit by random sampling and by agent uncertainty (\S\ref{sec:expert-loop}) before discarding and keep that designation while pending, are exempt from the discard rule and are served first through the audit allowance below.

The audit allowance $b^{\mathrm{sen}}_t$ is a fixed fraction of $b_t$ rather than a policy output. Serving pending sentinel events in order of capture time, the scheduler selects a set $\mathcal{P}^{\mathrm{sen}}_t$ whose planned payload $d^{\mathrm{sen}}_t$ fits within the allowance:
\begin{equation}
  d^{\mathrm{sen}}_t = \sum_{i \in \mathcal{P}^{\mathrm{sen}}_t} w_i
    \le b^{\mathrm{sen}}_t .
  \label{eq:sched-audit}
\end{equation}
Sentinel events that do not fit can still enter the task-directed plan if they pass the threshold, and any sentinel event not delivered remains pending. The remaining budget goes to task-directed uploads:
\begin{equation}
  \mathcal{P}^{\mathrm{task}}_t = \mathrm{SelectTop}\big(
    b_t - d^{\mathrm{sen}}_t,\, \gamma_t;\;
    \mathcal{B}_t \setminus \mathcal{P}^{\mathrm{sen}}_t \big).
  \label{eq:sched-selecttop}
\end{equation}
$\mathrm{SelectTop}$ keeps the candidates with $s_i \ge \gamma_t$ and ranks them by the confidence-weighted score $s_i(1 - u_i)$, so that among events with similar scores, those assessed more confidently rank higher. Uncertain events still reach experts through the audit allowance, which targets high uncertainty (\S\ref{sec:expert-loop}). Breaking ties in favour of older events, $\mathrm{SelectTop}$ traverses the ranked list and adds each event whose size fits within the remaining budget, skipping those that do not, so that one large event cannot leave the budget unused.

The upload plan $\mathcal{P}_t = \mathcal{P}^{\mathrm{sen}}_t \cup \mathcal{P}^{\mathrm{task}}_t$ therefore contains each event at most once and has total payload $\sum_{i \in \mathcal{P}_t} w_i \le b_t$. Before transmission, the safety shield (\S\ref{sec:sched-shield}) maps it to the executable plan $\mathcal{P}^{\mathrm{safe}}_t = \mathrm{Shield}(\mathcal{P}_t, \mathbf{o}_t)$, which is either $\mathcal{P}_t$ itself or the empty plan. Ordinary events that pass the threshold but are not successfully uploaded remain pending and are compared against the threshold of the next step.

\paragraph{Transition.} After the policy proposes $a_t$, the scheduler applies the discard rule to $\mathcal{B}_t$, constructs $\mathcal{P}_t$, and executes $\mathcal{P}^{\mathrm{safe}}_t$. The next state, with buffer $\mathcal{B}_{t+1}$ and energy $E_{t+1}$, reflects discarded events, completed uploads, new arrivals, realized harvest, and device load, with energy accounting covering both task-directed and sentinel traffic and the associated radio-state transitions. Delivered and discarded events leave the buffer. Retained events keep their capture times and continue to age, including through a shutdown, whereas observations missed during a shutdown are lost coverage and do not enter $\bar{L}_t$.

\paragraph{Satellite power-state control.} ATLAS switches the terminal between sleep and wake through Starlink's local gRPC interface, according to the executable plan and connectivity checks. Our measurements of the Starlink Standard (Gen 3) UT show power draws of 26.3\,W in sleep, 49.3\,W while idle, and 66.4\,W during active transmission, with a peak of 127.6\,W, and waking to a usable link takes 183\,s at 50.6\,W, or approximately 9.26\,kJ. Because the terminal returns to sleep at the end of each decision interval, every nonempty executed plan incurs this wake-up cost, which the scheduler accounts for with the idle and active energy that follows.

\begin{algorithm}[t]
\caption{Constrained SAC for ATLAS Upload Control}
\label{alg:csac}
\begin{algorithmic}[1]
  \State Initialize actor $\pi_{\phi}$ and twin critics $Q_{\psi_1}, Q_{\psi_2}$
  \State Initialize target critics, $\boldsymbol{\lambda} \ge 0$, and replay buffer $\mathcal{D}$
  \For{each multi-day training trace}
    \State Initialize battery and event-buffer state
    \For{each decision step $t$}
      \Statex \textit{Upload control}
      \State Observe $\mathbf{o}_t$; sample $a_t \sim \pi_{\phi}(\cdot \mid \mathbf{o}_t)$
      \State Update sentinel designations in $\mathcal{B}_t$
      \State Discard non-sentinel events with $s_i < \gamma_t$
      \State Build $\mathcal{P}_t = \mathcal{P}^{\mathrm{sen}}_t \cup \mathcal{P}^{\mathrm{task}}_t$ by Eqs.~(\ref{eq:sched-audit}) and~(\ref{eq:sched-selecttop})
      \State $\mathcal{P}^{\mathrm{safe}}_t \leftarrow \mathrm{Shield}(\mathcal{P}_t, \mathbf{o}_t)$
      \State Execute $\mathcal{P}^{\mathrm{safe}}_t$ and observe $\mathbf{o}_{t+1}$
      \State Compute $J_t$, $\mathbf{C}_t$, and terminal indicator $m_t$
      \State Append the transition to $\mathcal{D}$
      \Statex \textit{Policy learning}
      \State Sample a minibatch from $\mathcal{D}$
      \State Compute minibatch mean cost $\bar{\mathbf{C}}$
      \State Compute penalized rewards by Eq.~(\ref{eq:sched-penalized})
      \State Update twin critics on the penalized rewards
      \State Update $\pi_{\phi}$ using the soft-critic objective
      \State $\boldsymbol{\lambda} \leftarrow [\boldsymbol{\lambda} + \eta_{\lambda} (\bar{\mathbf{C}} - \boldsymbol{\delta})]_{+}$
      \State Soft-update target critics
    \EndFor
  \EndFor
\end{algorithmic}
\end{algorithm}

\subsection{Reward and Constraints}
\label{sec:sched-reward}

\paragraph{Delivered semantic value.} Let $\mathcal{U}_t \subseteq \mathcal{P}^{\mathrm{safe}}_t$ be the set of events successfully uploaded during step $t$. Their delivered semantic value is
\begin{equation}
  V_t = \sum_{i \in \mathcal{U}_t} s_i\,(1 - u_i),
  \label{eq:sched-value}
\end{equation}
the confidence-weighted score that $\mathrm{SelectTop}$ ranks by. Since $s_i$ estimates the probability that an expert would judge event $i$ to satisfy the intent and $u_i$ predicts the error of that estimate (\S\ref{sec:agent-design}), $V_t$ is a confidence-weighted count of the delivered clips expected to satisfy the intent.

\paragraph{Reward.} The per-step reward $J_t$ credits delivered value, charges data lag, and credits battery energy above the reserve while charging any shortfall below it:
\begin{equation}
  J_t = \omega_V V_t - \omega_L \bar{L}_t
        - \omega_E \big( E_{\min} - E_t \big),
  \label{eq:sched-reward}
\end{equation}
with nonnegative weights $\omega_V$, $\omega_L$, and $\omega_E$. Here $V_t$ reflects the executed plan, whereas $\bar{L}_t$ and $E_t$ describe the start of step $t$, so the effect of an action on lag and energy appears in the reward and costs of later steps.

\paragraph{Constraints.} Two constraint costs measure excess lag and energy-reserve shortfall:
\begin{equation}
  \begin{split}
    C^{(1)}_t &= \max\big(0,\, \bar{L}_t - L^{\mathrm{tgt}}\big), \\
    C^{(2)}_t &= \max\big(0,\, E_{\min} - E_t\big),
  \end{split}
  \label{eq:sched-costs}
\end{equation}
where $L^{\mathrm{tgt}}$ is the data-lag target. With $N \ge 1$ denoting the number of decision steps in an episode, the scheduler targets the constrained objective
\begin{equation}
  \begin{split}
    \max_{\pi_{\phi}}\; &\mathbb{E}_{\pi_{\phi}}
      \Big[ \sum_{t=0}^{N-1} \beta^{t} J_t \Big] \\
    \text{s.t.}\; &\frac{\mathbb{E}_{\pi_{\phi}}\big[ \sum_{t=0}^{N-1} C^{(j)}_t \big]}{\mathbb{E}_{\pi_{\phi}}[N]} \le \delta_j,
      \quad j \in \{1,2\},
  \end{split}
  \label{eq:sched-cmdp}
\end{equation}
where ${\beta \in [0,1)}$ is the discount factor, $\delta_j \ge 0$ is the budget on the mean per-step cost of constraint $j$, and expectations are taken over episode trajectories under $\pi_{\phi}$. Each constraint weights every visited decision step equally across episodes and reduces to the expected per-episode mean cost when all episodes have the same length.

\subsection{Policy Learning}
\label{sec:sched-learning}
We train the scheduler with a Lagrangian-style Soft Actor-Critic (SAC) procedure~\cite{haarnoja2018soft, ha2020learning, yang2021wcsac}, following reward constrained policy optimization~\cite{tessler2018reward}. The constraint costs of Eq.~(\ref{eq:sched-costs}) are folded into a penalized reward
\begin{equation}
  \tilde{J}_t = J_t - \boldsymbol{\lambda}^{\!\top} \mathbf{C}_t, \qquad
  \mathbf{C}_t = \big[ C^{(1)}_t,\, C^{(2)}_t \big]^{\!\top},
  \label{eq:sched-penalized}
\end{equation}
where $\boldsymbol{\lambda} = (\lambda_1, \lambda_2)^{\!\top} \ge 0$ are the Lagrange multipliers. SAC trains the actor $\pi_{\phi}$ and twin critics $Q_{\psi_1}$ and $Q_{\psi_2}$, with parameters $\psi_1$ and $\psi_2$, to maximize the entropy-regularized discounted return of $\tilde{J}_t$, so the penalty acts on discounted costs and steers the policy toward the mean-cost constraints of Eq.~(\ref{eq:sched-cmdp}) rather than encoding them exactly. The multipliers are updated by projected dual ascent~\cite{tessler2018reward} at rate $\eta_{\lambda} > 0$, rising whenever the minibatch mean constraint cost $\bar{\mathbf{C}}$, an off-policy proxy for the mean per-step cost in Eq.~(\ref{eq:sched-cmdp}), exceeds its budget $\boldsymbol{\delta} = (\delta_1, \delta_2)^{\!\top}$. Algorithm~\ref{alg:csac} summarizes the procedure, where $[\cdot]_{+}$ replaces negative components with zero and leaves nonnegative components unchanged. Three aspects are specific to \textsc{ATLAS}.

\paragraph{Continuous action.} SAC optimizes both outputs of $\pi_{\phi}$ directly, and the events to upload follow deterministically from them through plan construction (\S\ref{sec:sched-sat}).

\paragraph{Learning under the safety shield.} Each record in the replay buffer $\mathcal{D}$ stores the observation $\mathbf{o}_t$, the proposed action $a_t$, the reward $J_t$, the costs $\mathbf{C}_t$, the next observation $\mathbf{o}_{t+1}$, and an episode-termination indicator $m_t \in \{0,1\}$, equal to one if the episode ends after step $t$. The record keeps the proposed action rather than the executed plan $\mathcal{P}^{\mathrm{safe}}_t$, while $V_t$ and $\mathbf{o}_{t+1}$ reflect score-based discarding and any intervention by the shield (\S\ref{sec:sched-shield}), so the policy learns the consequences of its proposals as they are actually executed. Storing $J_t$ and $\mathbf{C}_t$ separately lets every update apply the current multipliers.

\paragraph{Training data and deployment.} Training uses multi-day trace-driven episodes built from deployment records of solar harvest, device load, uplink throughput and outages, and event arrivals. During replay, energy and buffer states evolve under the executed upload plans and the device-load accounting, exposing the policy to extended periods of resource scarcity. We initialize training with experience from historical operation and baseline-derived threshold policies, then optimize the scheduler off-policy. During deployment, $\pi_{\phi}$ is refined with conservative exploration under the same shield, while the agent's weights stay fixed (\S\ref{sec:agent-training}). This refinement is separate from the expert-feedback workflow, which adapts semantic assessment through the rubric and the retrieval stores.

\subsection{Runtime Safety Shield}
\label{sec:sched-shield}

The constraints of Eq.~(\ref{eq:sched-cmdp}) are imposed only in expectation, and the policy keeps exploring after deployment, so ATLAS also enforces per-step limits that every executed plan must satisfy under the current forecasts. Before execution, the shield checks two guards and defers the plan if either fails:
\begin{enumerate}
  \item \textbf{Energy protection.} The projected battery energy at the end of the interval, $\widehat{E}^{\,\mathrm{post}}_{t+1}$, must not fall below $E_{\min}$. The projection combines $E_t$, the harvest forecast, and the estimated terminal and device energy for the plan, including the wake-up cost.
  \item \textbf{Connectivity feasibility.} The link must not be predicted to be unavailable, such as under severe weather conditions~\cite{fang2025streaming}.
\end{enumerate}
The shield therefore yields $\mathcal{P}^{\mathrm{safe}}_t \in \{\mathcal{P}_t, \emptyset\}$ for task-directed and sentinel traffic alike, and the check is repeated just before transmission begins. A deferred plan leaves the terminal asleep, and its events remain pending. Because the guards act on forecasts, a harvest shortfall can still drain the battery (\S\ref{sec:ablation}). Data lag is not enforced by the shield. It is addressed during learning through the mean-cost constraint on $C^{(1)}_t$ in Eq.~(\ref{eq:sched-cmdp}) and is influenced at execution time by the age tie-break of $\mathrm{SelectTop}$.

\section{Expert-in-the-Loop Workflow}
\label{sec:expert-loop}
Human experts review uploaded clips in the cloud, and their feedback returns to the edge as updates to the context the agent reads (\S\ref{sec:agent-training}). Such review is standard practice in fisheries monitoring: trained reviewers annotate and audit electronic-monitoring video~\cite{noaa_em_guidance}, and automatic salmon counters are validated by manual comparison with video~\cite{scot_counter_network}. ATLAS builds on this existing review rather than adding a new one, and turns each review into feedback for the agent.

\paragraph{Sentinel sampling.} Task-directed uploads include only clips that pass the score threshold, so reviewing them alone is unlikely to reveal relevant events assigned low scores. ATLAS therefore selects sentinel candidates independently of their relevance scores through two complementary mechanisms. Random candidates are sampled within each regime defined by $r(c)$ to provide review coverage across the score range and observed conditions. Uncertainty-flagged candidates have high $u$, indicating that the agent expects their scores to be less reliable. Both types are eligible for the audit allowance described in \S\ref{sec:sched-sat} regardless of score.

\paragraph{Expert review.} Experts review both task-directed and sentinel uploads on the cloud platform (\S\ref{sec:eval}). For each clip, an expert records the judgment $z$ of whether it satisfies the current intent and marks the frames $\mathcal{A}(c)$ that decide it (\S\ref{sec:agent-training}). Experts can also add short comments on discriminative cues or failure causes, such as ``tail spots visible'' or ``glare/occlusion.''

\paragraph{Recurring feedback.} Feedback from our collaborating partner falls into four categories.
\begin{itemize}[leftmargin=1.5em, labelwidth=0.7em, labelsep=0.3em]
\item \textbf{Fine-grained visual discriminators.} Experts annotate cues such as tail and back markings, gumline colour, and body coloration. These cues extend the rubric's \texttt{evidence} field, so that the VLM reports them in every subsequent frame description.
\item \textbf{Counting under occlusion and bursty passage.} Experts explain how to count when tracks overlap or break during dense passage, and when a re-entering track belongs to an individual already counted. These rules are added to $\mathcal{K}$ and applied to the counts and track identities in $\mathrm{Describe}(c)$.
\item \textbf{Contextual false-positive signatures.} Comments such as ``debris drift,'' ``air bubbles,'' ``glare band,'' and ``turbidity plume'' are stored with the rejected clips as negative exemplars in $\mathcal{M}$, and recurring ones are added to $\mathcal{K}$ as known failure modes, so that similar observations receive lower scores.
\item \textbf{Intent-based relevance.} Guidance on first arrivals, rare species, and unusual behaviour helps the agent judge whether an observation adds evidence for the current intent or repeats a routine event. Such guidance is added to $\mathcal{K}$ as seasonal priors and rules of thumb, and, when it refines the intent itself, enters the rubric's \texttt{criteria} list at recompilation.
\end{itemize}

\paragraph{Synchronization.} Expert updates are batched in the cloud and delivered to the edge during the next upload window, so that synchronization never wakes the terminal on its own. A recompiled rubric is the only update that affects uploads directly, since the representation $\kappa$ is chosen from it (\S\ref{sec:sched-sat}). All other updates act only indirectly, by changing the scores $s$ and uncertainty estimates $u$ that the agent assigns to subsequent events.

%% file: sections/4-evaluation.tex
\section{Evaluation Methodology}
\label{sec:eval}

\begin{figure}
    \centering
    \includegraphics[width=\linewidth]{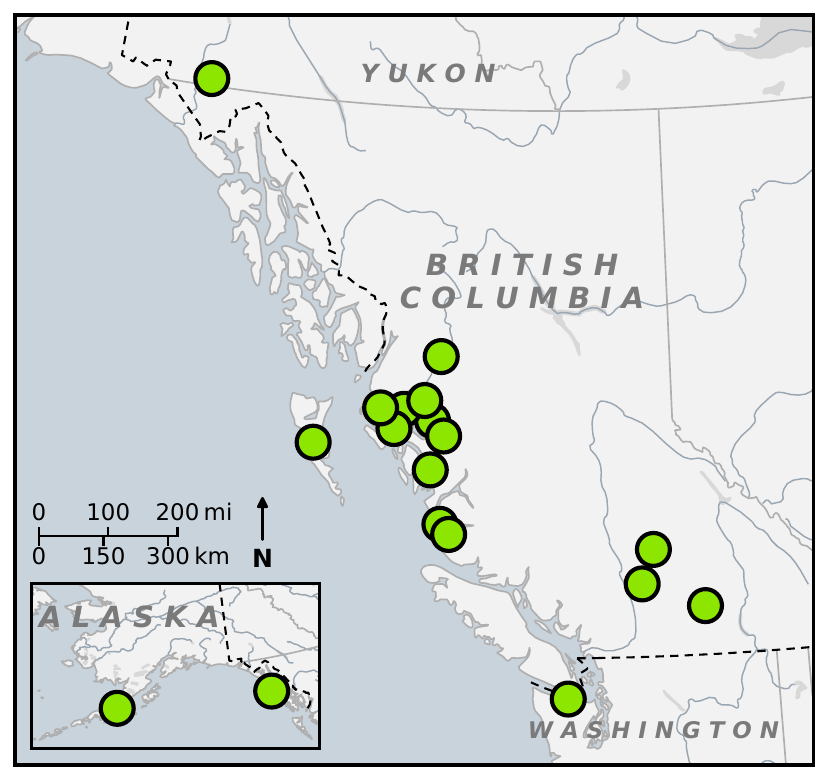}
    \vspace{-9mm}
    \caption{Geographic overview of ATLAS deployment sites operated in collaboration with our salmon fisheries partner across North America.}
    \vspace{-4mm}
    \label{fig:salmon_vision_deployment}
\end{figure}

\paragraph{Deployment and data.} We deployed ATLAS at 18 off-grid salmon monitoring sites operated with our fisheries partner (Figure~\ref{fig:salmon_vision_deployment}) from April to November 2025, spanning the full salmon migration and spawning season. Each site ran the hardware stack shown in Figure~\ref{fig:site_access}d. Throughout the deployment, every site recorded solar irradiance, weather, device load, uplink bandwidth, and camera footage containing both events and non-events. These recordings, together with the clips the partner had annotated before the deployment and its written domain knowledge, are the data behind every result in this paper. ATLAS ran live at the sites throughout, with experts reviewing uploads on the partner's platform hosted on an Amazon Web Services server. Each site has a single terminal and battery, and its conditions never repeat, so competing systems cannot be compared in the field. All head-to-head comparisons therefore replay the recorded traces on the deployment hardware, so that every system faces identical conditions.

\paragraph{Trace-driven replay.} Replay runs on an NVIDIA Jetson Orin NX~\cite{jetson_orin_nx} with 16~GB of shared CPU-GPU memory and 1~TB of local storage for video buffering, the same device as at the sites. ATLAS runs the entire on-device pipeline of Figure~\ref{fig:system_overview} on the Jetson. The recorded footage supplies the candidate events, and the recorded irradiance, weather, and bandwidth traces drive the simulated battery and link. Because the tasks require different event types and temporal contexts, no single recording window suits all of them. For each site and task, we therefore sample replay traces from the site's recordings suited to that task and fix them, so that ATLAS and every baseline run on the same traces, battery capacity, initial state of charge, and evaluation window, while each system's own execution determines its energy consumption, battery trajectory, and uptime. Replay traces are disjoint in time from the traces used to train ATLAS's scheduler and the Joint-optimized policy and from the footage used to fine-tune the DVM.

\paragraph{Implementation.} The DVM is a YOLO26s detector~\cite{yolo26} fine-tuned on annotated footage from the deployment sites, with ByteTrack~\cite{zhang2022bytetrack} for tracking, and the VLM is FastVLM-1.5B~\cite{fastvlm}. The agent's backbone is Qwen3-8B~\cite{yang2025qwen3} in non-thinking mode, fine-tuned with LoRA~\cite{hu2022lora} of rank 32 and quantized to 4-bit weights with AWQ \cite{lin2024awq}, so that it fits in the Jetson's shared memory alongside the DVM and the VLM \cite{tian2025clone}. Qwen3-Embedding-0.6B~\cite{zhang2025qwen3embedding} embeds the entries of $\mathcal{K}$ and $\mathcal{M}$ and the retrieval queries. Distillation used GPT-4o~\cite{gpt4o_2024-11-20} as the teacher, and refinement used 1,000 expert-reviewed clips disjoint from the evaluation traces, with $G = 8$ samples per input at temperature 1.0, $\rho = 0.5$, $\nu = 1$, $\epsilon = 10^{-3}$, and a KL coefficient of 0.03. Both stages ran on a server with four NVIDIA RTX A5000 GPUs.

\paragraph{Baselines.} 
We compare against two content-selection strategies that existing systems rely on, motion triggering and on-device vision, and against a baseline that adds learned energy-aware scheduling. The first two baselines upload the selected clips whenever the battery state of charge exceeds the reserve $\mathrm{SOC}_{\min}$ (\S\ref{sec:sched-sat}), the same reserve that ATLAS's safety shield protects, and the third replaces this rule with a learned policy. For each task, the experts specify the target classes and tune per-site confidence thresholds for the detector-based baselines, the closest approximation of the task that these systems can take as input. The ablation in \S\ref{sec:ablation} separates the contribution of each ATLAS component.
\begin{itemize}[leftmargin=1.5em, labelwidth=0.7em, labelsep=0.3em]
\item \textbf{Motion-based.} Motivated by on-camera filtering~\cite{li2020reducto}, this baseline selects clips by motion, using background subtraction with a mixture-of-Gaussians model.
\item \textbf{Vision-based.} Motivated by InFi~\cite{yuan2022infi}, this baseline selects clips with the same DVM as ATLAS, by the expert-specified classes and thresholds.
\item \textbf{Joint-optimized.} Inspired by learned video-quality adaptation~\cite{mao2017neural,sun2024biswift} and energy-aware transmission scheduling~\cite{schulman2010bartendr}, this baseline pairs the same DVM with an RL policy trained offline on the same deployment traces as ATLAS's scheduler. The policy takes the detection confidence as the relevance score, together with the battery state of charge, available bandwidth, and queue state, and decides whether to upload each clip and at what quality.
\end{itemize}

\begin{figure*}[!t]
    \centering
    \includegraphics[width=\linewidth]{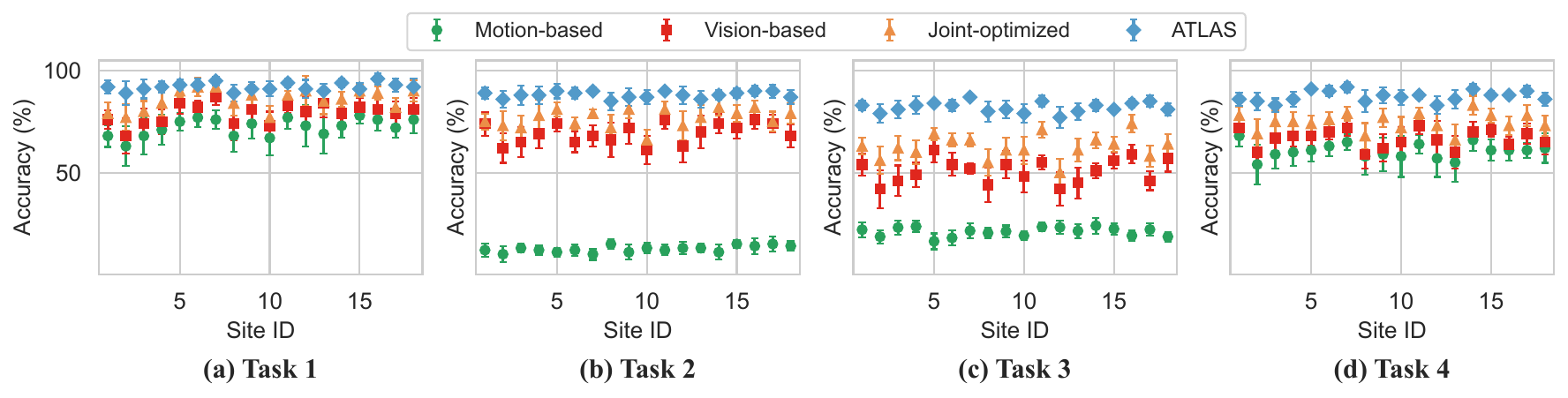}
    \vspace{-10mm}
    \caption{Clip-selection accuracy across sites for different tasks.}
    \label{fig:accuracy_sites}
    \vspace{-5mm}
\end{figure*}

\begin{figure*}[!t]
    \centering
    \includegraphics[width=\linewidth]{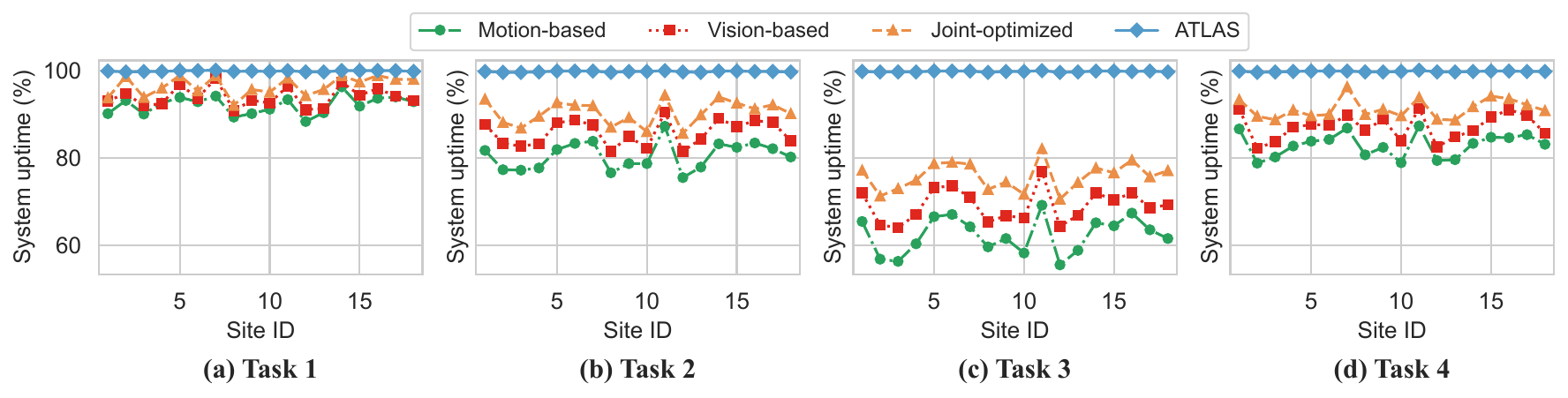}
    \vspace{-10mm}
    \caption{System uptime across sites for different tasks.}
    \label{fig:uptime_sites}
    \vspace{-5mm}
\end{figure*}

\begin{table}[t]
\centering
\caption{ATLAS versus three baselines across 18 deployment sites under four expert tasks. Accuracy is the percentage of candidate clips whose upload decision agrees with the expert; uptime is the percentage of the evaluation window in which the device is not shut down by the battery cutoff.}
\vspace{-4mm}
\label{tab:overall-results}
\small
\setlength{\tabcolsep}{3.2pt}
\begin{tabular*}{\columnwidth}{@{}l@{\extracolsep{\fill}}cccc@{}}
\toprule
\textbf{Metric / System} & \textbf{Task 1} & \textbf{Task 2} & \textbf{Task 3} & \textbf{Task 4}\\
\midrule
\multicolumn{5}{@{}l}{\textbf{\textit{Accuracy (\%)}}} \\
Motion-based    & 72.3 & 13.0 & 21.0 & 60.7 \\
Vision-based    & 79.1 & 69.4 & 50.8 & 66.7 \\
Joint-optimized & 85.7 & 76.6 & 62.6 & 74.8 \\
ATLAS & \textbf{92.1} & \textbf{88.2} & \textbf{82.0} & \textbf{87.4} \\
\midrule
\multicolumn{5}{@{}l}{\textbf{\textit{Uptime (\%)}}} \\
Motion-based    & 92.1 & 80.5 & 62.4 & 82.9 \\
Vision-based    & 93.9 & 85.7 & 69.1 & 87.1 \\
Joint-optimized & 96.5 & 90.3 & 75.9 & 91.2 \\
ATLAS & \textbf{99.9} & \textbf{99.7} & \textbf{99.9} & \textbf{99.7} \\
\bottomrule
\end{tabular*}
\vspace{-6mm}
\end{table}

\paragraph{Expert tasks.} To evaluate ATLAS under realistic expert-driven workflows, we worked with our partners to define four expert tasks drawn from real-world salmon monitoring deployments, as detailed below.

\emph{Task 1: Target Species Detection.} Identify and upload video clips that contain a specific target salmon species (e.g., Chinook salmon).

\emph{Task 2: Non-Target and Obstruction Identification.} Identify and upload clips containing non-target species or obstructions that may interfere with salmon passage observations.

\emph{Task 3: Passage Activity Monitoring.} Identify and upload video clips that help experts assess changes in salmon passage activity throughout the day, including increases, periods of high salmon passage activity, and decreases.

\emph{Task 4: Directional Migration Filtering.} Upload video clips that capture upstream salmon passage while filtering downstream movement and non-migratory behaviour.

\paragraph{Metrics.} Clip-selection accuracy is the percentage of candidate clips whose outcome, uploaded within the evaluation window or not, agrees with the expert's judgment $z$, so it penalizes both irrelevant uploads and missed relevant clips. Every candidate clip in the replay set carries such a judgment, collected independently of the clips used for refinement (\S\ref{sec:agent-training}), and sentinel uploads are counted like any other. \emph{System uptime} is the percentage of the evaluation window in which the edge device is not shut down by the battery cutoff. Both are reported per site and task. \emph{Data lag}, the mean age $\bar{L}_t$ of pending events (\S\ref{sec:sched-sat}), is shown over time in \S\ref{sec:seasonal}.

\section{Performance Evaluation}
\label{sec:perf}

\subsection{Overall Performance}
Table~\ref{tab:overall-results} summarizes the overall performance of ATLAS compared with three representative baselines across 18 real-world deployments and four expert-defined tasks. We evaluate performance along two key dimensions that are important for long-term remote monitoring: system uptime and clip-selection accuracy.

\textbf{System uptime.} ATLAS maintains near-continuous operation, with task-level uptime ranging from $99.7\%$ to $99.9\%$ across the four evaluated workloads. In contrast, baseline systems exhibit substantial downtime under energy-constrained conditions, with motion-based approaches degrading to as low as 62.4\% uptime in Task~3. Even the joint-optimized baseline, which explicitly accounts for system constraints, experiences non-trivial outages (up to 24.1\%). These results indicate that static or offline-optimized policies are insufficient to handle prolonged low-energy periods and task-dependent workload variation, whereas ATLAS proactively adapts its uplink behaviour to both its energy state and expert intent.

\textbf{Clip-selection accuracy.} ATLAS achieves the highest accuracy on all four tasks, exceeding the strongest baseline on each by 6.4 to 19.4 percentage points. The margin is largest on Task 3, where ATLAS reaches 82.0\%, compared with 62.6\% for Joint-optimized and 21.0\% for Motion-based. Averaged equally across tasks, ATLAS achieves a mean accuracy of 87.4\%, 26.4 percentage points above the baseline mean. Together with the uptime results, these findings show that ATLAS improves accuracy while maintaining near-continuous operation under the evaluated energy constraints.

\begin{figure}[!t]
    \centering
    \includegraphics[width=\linewidth]{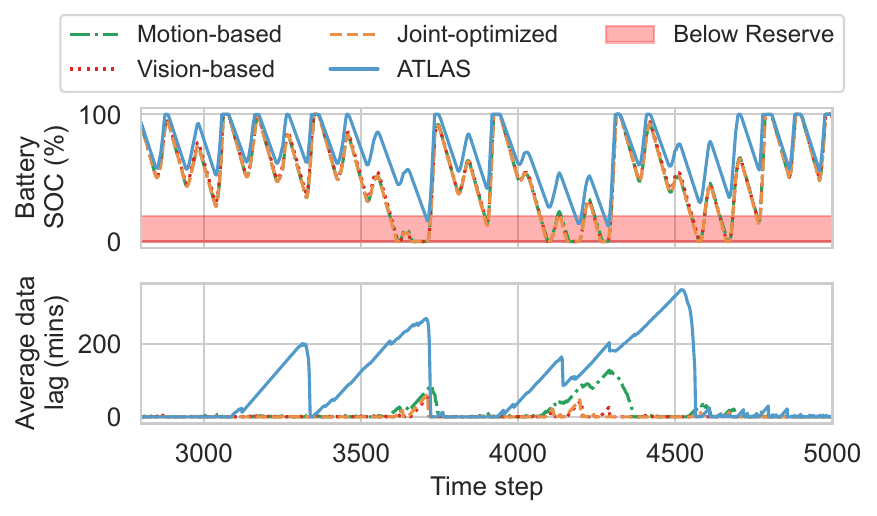}
    \vspace{-7mm}
    \caption{Battery SOC (top) and average data lag (bottom) of ATLAS and the baselines during a prolonged rainy period.}
    \vspace{-4mm}
    \label{fig:show_case}
\end{figure}

\subsection{Performance Across Sites and Tasks}
\textbf{Accuracy.} Figure~\ref{fig:accuracy_sites} shows the per-site accuracy behind the averages in Table~\ref{tab:overall-results}, from the same replay traces (\S\ref{sec:eval}). ATLAS achieves the highest accuracy at every site and on every task, with low variance across sites. Its margin over the baselines is smallest on Task~1 but consistent across sites. The advantage is largest on Task~3, which requires clips that help experts assess changes in passage activity throughout the day: baseline accuracy drops sharply and varies widely across sites, while ATLAS remains stable, showing robustness to complex temporal and contextual requirements.

\textbf{System uptime.} Figure~\ref{fig:uptime_sites} presents the site-level uptime results for the four task-specific replay workloads. The corresponding task-level aggregates in Table~\ref{tab:overall-results} range from $99.7\%$ to $99.9\%$ for ATLAS. These results demonstrate near-continuous operation under the evaluated conditions, alongside the improvements in clip-selection accuracy.

\textbf{Summary.} Across heterogeneous sites and increasingly complex tasks, ATLAS delivers consistently high accuracy while maintaining near-continuous operation.

\subsection{Performance During Worst-Case Seasonal Conditions}
\label{sec:seasonal}
Figure \ref{fig:show_case} illustrates system behaviour during one of the most challenging seasonal periods, characterized by prolonged rainfall and limited solar input. Under these conditions, battery SOC falls repeatedly below the reserve and down to the cutoff across all baselines, causing frequent system shutdowns. In contrast, ATLAS maintains battery SOC above the reserve throughout the period by proactively budgeting energy and deferring video uploads. This energy stability comes at the cost of increasing data lag as deferred uploads accumulate during sustained energy scarcity. Once energy availability recovers, ATLAS clears the accumulated upload backlog, returning data lag to near zero while preserving uninterrupted system operation.

\begin{table}[t]
\centering
\caption{Ablation results for ATLAS.}
\vspace{-4mm}
\label{tab:ablation-main}
\small
\setlength{\tabcolsep}{3.2pt}
\begin{tabular*}{\columnwidth}{@{}l@{\extracolsep{\fill}}cccc@{}}
\toprule
\textbf{Metric / Variant} & \textbf{Task 1} & \textbf{Task 2} & \textbf{Task 3} & \textbf{Task 4}\\
\midrule
\multicolumn{5}{@{}l}{\textbf{\textit{Accuracy (\%)}}} \\
w/o upload scheduler          & 90.0 & 86.0 & 79.0 & 85.1 \\
w/o semantic assessment agent & 85.0 & 77.1 & 68.2 & 76.5 \\
w/o expert-in-the-loop        & 89.9 & 82.7 & 78.9 & 83.5 \\
ATLAS & \textbf{92.1} & \textbf{88.2} & \textbf{82.0} & \textbf{87.4} \\
\midrule
\multicolumn{5}{@{}l}{\textbf{\textit{Uptime (\%)}}} \\
w/o upload scheduler          & 93.2 & 92.1 & 87.0 & 94.1 \\
w/o semantic assessment agent & 99.2 & 98.4 & 96.8 & 98.9 \\
w/o expert-in-the-loop        & 94.1 & 93.2 & 95.4 & 97.5 \\
ATLAS & \textbf{99.9} & \textbf{99.7} & \textbf{99.9} & \textbf{99.7} \\
\bottomrule
\end{tabular*}
\vspace{-3mm}
\end{table}

\section{Ablation Study}
\label{sec:ablation}

We evaluate three ablation variants, each removing one ATLAS component. \emph{W/o upload scheduler} keeps the semantic assessment agent and expert updates but replaces the scheduler with a first-come, first-served policy: clips that pass a fixed score threshold are uploaded in arrival order whenever the battery state of charge is above the safety shield's reserve threshold. \emph{W/o semantic assessment agent} keeps the upload scheduler, safety shield, and expert updates, but scores each clip by the DVM’s confidence for the target classes and sets its uncertainty estimate to zero. Since this variant has no agent context to update, expert feedback instead tunes the DVM’s target classes and confidence thresholds. \emph{W/o expert-in-the-loop} removes all expert updates while preserving the rest of the ATLAS runtime.

Table~\ref{tab:ablation-main} reports accuracy and uptime on the four tasks. The full system is the most accurate on every task, with a mean accuracy of 87.4\%. Removing the semantic assessment agent causes the largest loss, lowering mean accuracy to 76.7\%, a drop of 10.7 percentage points that ranges from 7.1 to 13.8 points across tasks. The loss is largest on Task 3, whose temporal and contextual requirements cannot be expressed through DVM classes and thresholds, even with expert tuning. Removing expert-in-the-loop updates lowers mean accuracy to 83.8\%, a drop of 3.7 points, and removing the upload scheduler lowers it to 85.0\%, a drop of 2.4 points.

Removing the upload scheduler causes the largest uptime loss. Mean uptime falls from 99.8\% to 91.6\%, a decrease of 8.2 points, with the largest drop on Task 3. Without the scheduler's forward-looking energy budget, uploads continue whenever the battery is above the reserve threshold, so the battery is often close to that threshold when a low-harvest period begins.

Removing expert-in-the-loop updates reduces uptime by 2.2 to 6.5 points across tasks, with a mean decrease of 4.8 points, even with the scheduler and safety shield enabled. Without updated expert knowledge, irrelevant clips can receive high relevance scores, and uploading them leaves less energy for low-harvest periods. Because the energy guards rely on harvest predictions, a harvest shortfall can still exhaust the battery after uploads are suspended, so earlier upload decisions affect uptime even with the guards in place.

In contrast, removing the semantic assessment agent reduces uptime by only 0.7 to 3.1 points, with a mean decrease of 1.5 points. Expert-tuned DVM thresholds make this variant conservative: its accuracy loss comes mainly from relevant clips it fails to select, not from irrelevant clips it uploads, so it spends less energy on uploads.

Together, these results separate the roles of the three components. The semantic assessment agent contributes most to accuracy, the upload scheduler contributes most to uptime, and expert-in-the-loop updates improve both. Expert tuning of DVM thresholds does not recover the agent's accuracy, and no variant matches the full system on either metric.

%% file: sections/5-related_works.tex
\section{Related Work}
\textbf{Energy- and connectivity-constrained sensing.} Intermittent backhaul and reliance on harvested energy are long-standing constraints on unattended sensor deployments. Delay-tolerant architectures move data opportunistically rather than assuming a continuous end-to-end path~\cite{fall2003dtn,park2011datamuling,dyo2010wildlife,juang2002zebranet, xu2024salina}, operating systems manage energy as an explicit, allocatable resource~\cite{zeng2003currentcy}, energy-neutral systems pair harvesting with multi-stage storage for perpetual operation~\cite{jiang2005prometheus} and adapt duty cycles and application activity to predicted harvest~\cite{kansal2007power,sorber2007eon}, and transmission schedulers defer sending to moments when it is cheaper or more informative~\cite{schulman2010bartendr,wu2012event}. In wildlife monitoring, edge processing with cloud support shrinks the data that must cross constrained links~\cite{elias2017s}, and a recent proposal keeps expert knowledge in an explicit knowledge base on the edge so that adaptation does not require retraining~\cite{ijcai2026p811}. None of these systems weighs what a transmission is worth to the experts who use the data beyond a fixed proxy, a priority level, an object class, or a measure of information. \textsc{ATLAS} determines what each clip is worth at runtime from expert intent that changes over a deployment.

\textbf{Edge/on-camera video analytics and bandwidth reduction.} Systems research has developed rich mechanisms to reduce video analytics cost by dynamically trading accuracy for resource usage through configuration adaptation and runtime control~\cite{jiang2018chameleon,zhang2018awstream,zhang2017videostorm,hsieh2018focus}, and by filtering frames early, often close to the camera, to avoid processing or transmitting uninformative content~\cite{elias2017s,hsieh2018focus,li2020reducto,simoens2013gigasight,hwang2022cova}. Zero-streaming cameras go further, storing video locally and transferring it only when an analyst issues a query~\cite{xu2021video}. These approaches typically optimize generic accuracy and throughput objectives or respond to explicit queries, whereas \textsc{ATLAS} decides proactively what to upload based on \emph{expert intent} and co-optimizes this decision against energy-driven uptime constraints.

\textbf{LEO satellite networking.} 
Measurement, simulation, and emulation studies have characterized LEO network behaviour, from constellation-scale dynamics~\cite{kassing2020hypatia, fang2025streaming} and the energy, thermal, and network conditions of satellite computing~\cite{wang2025emulating} to the performance of Starlink's network and applications~\cite{mohan2024starlink, fang2024toward}. Other work advances satellite communication architectures. L2D2 uses distributed ground stations to reduce data-delivery latency from Earth-observation satellites~\cite{vasisht2021l2d2}. SateRIoT combines link estimation and inter-gateway data sharing to improve the energy efficiency and performance of satellite IoT backhaul~\cite{ren2024sateriot}, while MOSAIC supports multi-tenant direct-to-cell satellite access~\cite{liu2024directtocell}. At the application layer, StarStream adapts live video analytics to Starlink uplink variability through Transformer-based throughput prediction and content-aware configuration optimization~\cite{starstream}. \textsc{ATLAS} addresses the complementary problem of operating the terminal under a variable solar-energy budget, accounting for its wake-up and operating energy so that uploads are planned only when the site can afford them.

%% file: sections/6-conclusion.tex
\section{Conclusion}
We presented ATLAS, an expert-guided, energy-aware system that treats satellite backhaul as a scarce resource, combining FM-based interpretation of expert intent in natural language with constrained upload scheduling. In trace-driven comparisons on salmon monitoring data covering an entire migration and spawning season at 18 off-grid river sites, ATLAS achieved 87.4\% mean clip-selection accuracy across four expert-defined tasks, 26.4 percentage points above the baseline mean, with the largest gains on complex tasks. It also maintained near-continuous operation through low-solar periods, whereas baselines averaged 16.0\% downtime. More broadly, these results suggest that robust remote wildlife monitoring requires joint design of domain-grounded semantic reasoning and resource-aware control, not efficient edge inference or opportunistic connectivity alone.